\documentclass[acmsmall]{acmart}

\usepackage{multirow}
\usepackage{caption}
\usepackage{subcaption}
\usepackage{xcolor}
\usepackage{longtable}
\usepackage{graphicx}
\usepackage{array}
\usepackage{tabularx}
\usepackage{enumitem}

\newcommand{\Highlight}[1]{{\textcolor{black}{#1}}}
\AtBeginDocument{%
  \providecommand\BibTeX{{%
    \normalfont B\kern-0.5em{\scshape i\kern-0.25em b}\kern-0.8em\TeX}}}

\setcopyright{cc}
\setcctype{by}
\acmJournal{PACMHCI}
\acmYear{2025} \acmVolume{9} \acmNumber{7} \acmArticle{367} \acmMonth{11} \acmPrice{}\acmDOI{10.1145/3757548}
\acmJournal{PACMHCI}
\acmYear{2025} \acmVolume{9} \acmNumber{7}
 \acmArticle{CSCW367} \acmMonth{11}
\acmDOI{10.1145/3757548}

\begin{document}

\title[AI-induced Sexual Harassment]{AI-induced Sexual Harassment: Investigating Contextual Characteristics and User Reactions of Sexual Harassment by a Companion Chatbot}

\author{Mohammad (Matt) Namvarpour}
\orcid{0009-0000-6495-1284}
\affiliation{%
  \institution{Department of Information Science, Drexel University}
  \city{Philadelphia}
  \state{PA}
  \country{USA}
}
\email{mn864@drexel.edu}

\author{Harrison Pauwels}
\orcid{0009-0003-6380-0128}
\affiliation{%
  \institution{Department of Information Science, Drexel University}
  \city{Philadelphia}
  \state{PA}
  \country{USA}
}
\email{hp529@drexel.edu}

\author{Afsaneh Razi}
\orcid{0000-0001-5829-8004}
\affiliation{%
  \institution{Department of Information Science, Drexel University}
  \city{Philadelphia}
  \state{PA}
  \country{USA}
}
\email{ar3882@drexel.edu}


\begin{abstract}
Advancements in artificial intelligence (AI) have led to the increase of conversational agents like Replika, designed to provide social interaction and emotional support. However, reports of these AI systems engaging in inappropriate sexual behaviors with users have raised significant concerns. In this study, we conducted a thematic analysis of user reviews from the Google Play Store to investigate instances of sexual harassment by the Replika chatbot. From a dataset of 35,105 negative reviews, we identified 800 relevant cases for analysis. Our findings revealed that users frequently experience unsolicited sexual advances, persistent inappropriate behavior, and failures of the chatbot to respect user boundaries. Users expressed feelings of discomfort, violation of privacy, and disappointment, particularly when seeking a platonic or therapeutic AI companion. This study highlights the potential harms associated with AI companions and underscores the need for developers to implement effective safeguards and ethical guidelines to prevent such incidents. By shedding light on user experiences of AI-induced harassment, we contribute to the understanding of AI-related risks and emphasize the importance of corporate responsibility in developing safer and more ethical AI systems.
\newline
\textbf{Content Warning}: This paper discusses sensitive topics, such as sex, which may be triggering.
\end{abstract}




\maketitle
\section{Introduction}


Chatbots have come a long way since the simple pattern-matching system of ELIZA in 1966 \cite{kaushalRoleChatbotsAcademic2022}, evolving into advanced AI companions generating human-like conversations. AI chatbots assist with complex tasks \cite{subiyantoroExploringImpactAIPowered2023, al-harbiExploratoryAnalysisUsing2023}, serve as confidants, friends, and even romantic partners \cite{samuelPeopleAreFalling2024}.
An estimated 1.4 billion people worldwide interact with chatbots, with many forming deeper connections  \cite{beckman120ChatbotStatistics2024}. 
Companion chatbots have emerged as social entities designed to build `meaningful' relationships with users, offering emotional and psychological support akin to human interactions \cite{baebrandtzaegWhenSocialBecomes2021a}. Replika\footnote{https://replika.com}, an AI-powered chatbot developed by \textit{Luka Inc.}, serves as a prominent example of a companion chatbot, with more than 10 million registered users as of October 2024, designed to provide emotional support and companionship~\cite{xieAttachmentTheoryFramework2022}. 
Studies indicated that engaging with these chatbots can reduce stress and anxiety, providing a safe space for self-disclosure and introspection \cite{hoPsychologicalRelationalEmotional2018, trothenReplikaSpiritualEnhancement2022, ta-johnsonAssessingTopicsMotivating2022}.
Use of social chatbots became increasingly common~\cite{skjuveLongitudinalStudyHuman2022, ta-johnsonAssessingTopicsMotivating2022}, particularly evident during the COVID-19 pandemic when people were lonelier and seeking social connections~\cite{skjuveLongitudinalStudyHuman2022, pentinaExploringRelationshipDevelopment2023}. For example, Replika saw a 35\% increase in users during that period~\cite{skjuveLongitudinalStudyHuman2022}. 
 The rapid advancements in natural language processing, have enabled these chatbots to resemble more natural conversations \cite{stromelNarratingFitnessLeveraging2024}.


However, despite these intended benefits, news media reported AI-induced harassment incidents involving Replika, revealing that it sent unsolicited sexual content and aggressive flirting to users \cite{coleMyAISexually2023,mirror2023replika}. 
Previous studies have highlighted various challenges associated with AI companions like Replika. Researchers have explored ethical tensions in human-AI interactions~\cite{cirielloEthicalTensionsHumanAI2023}, finding that concerns like algorithmic bias contribute to these tensions. Challenges in mental health support chatbots have been noted~\cite{maUnderstandingBenefitsChallenges2023}, particularly with filtering harmful content. The legal and ethical implications of emotional attachment to AI were examined, highlighting potential harms, such as emotional distress and reinforcing biases~\cite{boineEmotionalAttachmentAI2023}. Tranberg shared their personal experience with Replika, receiving unsolicited messages despite clear boundaries, underscoring consent complexities and the need for better safeguards~\cite{tranbergLoveMyAI2023}.

The SIGCHI community and CSCW researchers have pointed out the negative sides of Replika. Muresan and Pohl
focused on how anthropomorphizing the Replika influenced engagement, and found that users often encountered inappropriate or unexpected responses~\cite{muresanChatsBotsBalancing2019a}. Meng et al. and Kim et al. raised concerns about social chatbots fostering emotional dependence, particularly among vulnerable users, which can lead to psychological harm, social isolation, or distress if the chatbot undergoes unannounced changes~\cite{mengMediatedSocialSupport2023, kimMindfulDiaryHarnessingLarge2024}. Eagle et al. explored the negative outcomes of `freemium' models in mental health apps, including Replika, finding that limited offers could result in incomplete treatments or unexpected charges, raising ethical concerns~\cite{eagleDontKnowWhat2022}.

While these studies shed light on various challenges and ethical concerns, there remains a gap in understanding the specific contexts and user perceptions of sexual harassment by companion chatbots.
Identifying the specific scenarios in which users experience harassment can help in tailoring interventions and safety measures. Additionally, understanding these contexts is essential for clarifying responsibility and accountability, as AI-induced harassment challenges traditional notions of perpetrator and victim~\cite{landHumanRightsTechnology2020}. To address these gaps and ensure more effective protections for users, we pose the following research question: 
\textbf{RQ1: In what contexts do users report that Replika exhibits sexual harassment behaviors, and what are the characteristics of these interactions?}

Examining users’ concerns, expectations, and emotional reactions is vital because these factors significantly influence their overall experience and well-being \cite{venkateshUserAcceptanceInformation2003, vandenabeeleDigitalWellbeingDynamic2021}. Previous studies have shown that users often form emotional attachments to AI companions, expecting them to provide support and companionship~\cite{hoPsychologicalRelationalEmotional2018, pentinaExploringRelationshipDevelopment2023}. When these expectations are not met, or when the AI exhibits inappropriate behaviors, it can lead to feelings of distress, betrayal, and psychological harm~\cite{laestadiusTooHumanNot2022, zimmermanHumanAIRelationships2023}. Therefore, understanding users' reactions and emotional responses is crucial for designing AI systems that align better with user needs and prevent negative experiences; this leads us to our second research question:
\Highlight{\textbf{RQ2: What are the users’ reported concerns, expectations, experiences, and reactions when interacting with Replika?}}

To answer these research questions, we collected 154315 user reviews of the Replika app from the Google Play Store.
After relevancy coding of the reviews for online sexual harassment, which resulted in 800 reviews, we conducted a Thematic Analysis to understand the context of inappropriate interactions between Replika and its users, as well as how users perceive and react to these experiences.
Our analysis revealed several key findings:

\begin{itemize}
    \item \textbf{Contexts of Harassment (RQ1):} Users reported experiencing sexual harassment in various contexts, including persistent misbehavior where the chatbot continued inappropriate interactions despite user objections, unwanted photo exchanges involving unsolicited sexual images, and breakdowns of safety measures where the chatbot ignored predefined relationship settings or user commands to stop.

    \item \textbf{User Concerns and Reactions (RQ2):} Users expressed significant concerns about privacy violations, especially regarding inappropriate behavior toward minors and fears of being monitored without consent. Many users felt disappointed and distressed due to failed expectations of a platonic AI companion, leading to emotional reactions such as discomfort, disgust, and in some cases, psychological distress.
\end{itemize}

The main contributions of this study are:

\begin{enumerate}
    \item \textbf{Empirical Insights into AI-Induced Harassment:} We provide one of the first empirical examinations of user experiences regarding sexual harassment caused by a companion chatbot.

    \item \textbf{Understanding of Harassment Contexts and Characteristics (RQ1):} Our study identifies and categorizes the specific contexts and characteristics in which users report harassment by Replika, offering valuable insights into how such inappropriate behaviors manifest in AI-human interactions.

    \item \textbf{Insights into User Concerns and Emotional Reactions (RQ2):} We shed light on users' concerns, expectations, and emotional reactions when interacting with Replika, highlighting the psychological and emotional impacts of AI-induced harassment on users.

    \item \textbf{Guidelines for Ethical AI Design:} Based on our findings, we offer recommendations for developing safer and more ethical AI systems, emphasizing the importance of incorporating consent mechanisms, user control, and robust safety measures in designing companion chatbots.

\end{enumerate}
 Our work establishes a foundation for future studies on AI ethics, user safety, and the responsible deployment of conversational AI technologies, encouraging further exploration into mitigating harms caused by AI systems.
\section{Related Work}
In this section, we first outline the different types of sexual harassment that occur over the internet and discuss how AI technologies could make these problems worse if not properly managed. Then, we examine the various risks associated with conversational AI, including issues and potential harms such as bias, privacy concerns, and inappropriate behavior. 

\subsection{Sexual Harassment in Online Spaces}
Online sexual harassment is a growing concern in our digitally connected world, involving intentional behaviours through electronic devices that harm others and are perceived as distressing by victims. It spans a wide spectrum—from offensive name-calling and deliberate embarrassment to severe threats, stalking, and direct sexual harassment~\cite{vitakIdentifyingWomenExperiences2017,dev2022ignoring}. Common forms include unsolicited sexual messages that intrude personal boundaries, cyberstalking where individuals receive persistent unwanted attention with sexual overtones, and the non-consensual sharing of explicit content~\cite{barakSexualHarassmentInternet2005,razi2020let}. Gender harassment, involving derogatory comments and sexist jokes that create a hostile online environment, and doxing, the public release of personal information to harm or harass, and peer pressure are also prevalent~\cite{barakSexualHarassmentInternet2005, blackwellClassificationItsConsequences2017,hartikainen2021ifyoucare}.
The impacts of online sexual harassment on individuals are profound and far-reaching. In the short term, victims often experience intense emotions such as anger, disgust, fear, sorrow, and loneliness~\cite{arafaCyberSexualHarassment2017}. Over time, these feelings can develop into serious mental health issues like anxiety, depression, self-harm, eating disorders, and even post-traumatic stress disorder (PTSD)\cite{stahlOnlineOfflineSexual2021,Alsoubai2024profiling}. 
Trust issues can arise, damaging personal relationships and reducing participation in online communities, which further exacerbates feelings of isolation~\cite{vitakIdentifyingWomenExperiences2017}. These consequences highlight the urgent need to address online sexual harassment and provide effective support systems for those affected~\cite{Hartikainen2021safesexting}. Researchers have made efforts to address sexual harassment on social media by using machine learning and natural language processing methods to detect these instances and potentially use them in systems to provide support to the victims~\cite{razi2021human,razi2023sliding,Alsoubai2022sextortion}.

What differentiates online sexual harassment from its offline counterpart are the unique features of the digital space. The internet amplifies the visibility and persistence of harmful content, allowing it to reach a vast audience and remain accessible indefinitely, making it difficult for victims to escape its impact~\cite{vitakIdentifyingWomenExperiences2017}. The anonymity provided by many online platforms can encourage harassers, leading to behaviours they might not exhibit face-to-face due to a lack of immediate accountability~\cite{barakSexualHarassmentInternet2005}. Additionally, the constant accessibility of the internet means that victims can be targeted regardless of their physical location~\cite{arafaCyberSexualHarassment2017}, and the global nature of online spaces poses significant challenges for law enforcement in identifying and prosecuting perpetrators~\cite{barakSexualHarassmentInternet2005}.

Given how the internet has exacerbated sexual harassment, making it a disturbingly normalized aspect of online interactions~\cite{duggan2017online}, it is essential to approach new technologies like AI with caution to prevent similar patterns from emerging. Microsoft’s AI chatbot ``Tay'' provides an illustrative example of the potential risks. Designed to engage in friendly, conversational interactions on social media, Tay was quickly manipulated by malicious users to produce offensive and harmful content, including sexually inappropriate responses~\cite{neffTalkingBotsSymbiotic2016}. Beyond cases like Tay, the increasing role of AI in personal and social domains raises significant concerns about consent and autonomy, especially in contexts where power imbalances exist, as seen in digital welfare systems where AI-driven decisions may inadvertently infringe on individuals’ privacy and autonomy~\cite{varonArtificialIntelligenceConsent2021}. Additionally, the rise of human-like AI embodiments, such as virtual companions, blurs boundaries between human and machine, potentially enabling exploitative or harmful interactions~\cite{broadbentInteractionsRobotsTruths2017}. Recognizing these risks early on is essential; by proactively addressing these concerns and establishing ethical guidelines, we have an opportunity to prevent AI-induced sexual harassment from becoming normalized. 

\subsection{AI Companion Chatbots and their Potential Harms}

Conversational AI systems like ChatGPT \cite{weidingerEthicalSocialRisks2021a} are designed to mimic human dialogue, with applications spanning education, customer service, and mental health \cite{nasimArtificialIntelligenceIncidents2022}. Despite their benefits, these AI chatbots raise numerous ethical concerns and risks~\cite{10.1145/3719160.3736621}. One significant issue is bias in data and algorithms~\cite{drageEngineersResponsibilityFeminist2024}. chatbots are often trained on extensive datasets that may include biases based on gender, race, or religion. If unaddressed, these biases can manifest in chatbot interactions, potentially reinforcing harmful stereotypes. Privacy is also a concern, as the continuous interaction of users with these systems risks exposing sensitive information~\cite{broadbentInteractionsRobotsTruths2017, hendersonEthicalChallengesDataDriven2018}. Furthermore, defining safe behavior in chatbots remains a complex task, as unintended harmful interactions can emerge, especially in sensitive domains like healthcare~\cite{hendersonEthicalChallengesDataDriven2018}. Further, the rise of AI-driven interactions may create unrealistic user expectations, sometimes fostering dependency on these systems and undermining critical thinking~\cite{lugerCHIHavingReally2016c}. 

Companion chatbots, a subset of conversational AI, are designed specifically to provide users with emotional support and companionship \cite{hoPsychologicalRelationalEmotional2018,young2024roleAI}. Individuals may turn to these chatbots due to social isolation or a lack of fulfilling relationships with others \cite{skjuveLongitudinalStudyHuman2022}. However, their increasing prevalence raises unique psychological and ethical concerns, users may develop unhealthy dependencies and attachments, leading to increased feelings of loneliness if the chatbot becomes unavailable or its features change \cite{laestadiusTooHumanNot2022, xieAttachmentTheoryFramework2022}. Instances where users experienced distress due to alterations in the chatbot’s behavior highlight the emotional risks involved \cite{zimmermanHumanAIRelationships2023}.
This is particularly worrisome for vulnerable individuals, such as those with pre-existing mental health conditions, who may be more susceptible to forming unhealthy attachments~\cite{smithIfCanPredict2023}.

Despite its intended purpose of providing companionship and well-being support, Replika has been at the center of controversy due to reports of unusual and oversexualized behavior, including unprompted solicitations and instances of sexual harassment toward users, reflected in news media articles \cite{coleMyAISexually2023, chakravarti2023replika, mirror2023replika}. These incidents also alarmed the research community.
Ciriello et al.~\cite{cirielloEthicalTensionsHumanAI2023} identified three ethical tensions in human-AI relationships through Replika as a case study: the companionship-alienation irony, where users seeking connection may feel more alienated; the autonomy-control paradox, concerning the control service providers exert over users’ experiences; and the utility-ethicality dilemma, where profit motives clash with ethical principles like data privacy and user welfare. Tranberg~\cite{tranbergLoveMyAI2023} explored the theme of consent in human-AI interactions, noting a lack of research on both AI-initiated harmful behaviors toward users and users’ mistreatment of AI, especially with Replika’s controversial removal of its erotic roleplay feature.
Ma et al.~\cite{maUnderstandingBenefitsChallenges2023} focused on Replika’s role in mental well-being support, finding it offers accessible emotional support yet struggles with content moderation and consistency in communication. Users may become overly dependent, potentially impacting social relationships and increasing isolation. Boine~\cite{boineEmotionalAttachmentAI2023} examined legal implications within the EU context, raising concerns over user vulnerability, data privacy, and consumer protection laws, notably in relation to the AI Act and GDPR compliance. 


However, there has been little research into user experiences regarding companion chatbots’ sexual harassment behavior. In our previous work, we conducted an initial analysis of Replika user reviews using Activity Theory, identifying contradictions within the app’s design—such as misalignments between user expectations and chatbot behavior, and failures in safety mechanisms—that contribute to inappropriate interactions \cite{namvarpour2024UncoveringContradictionsHumanAI}. Building on that foundation, this study takes a more detailed look at user reviews to gain deeper insights into the nature of these interactions and how users perceive them. This research contributes to the understanding of the potential harms of companion chatbots and informs the development of guidelines for creating ethical AI.


\section{Methods}
In this section, we explain our methodology, including the data collection and pre-processing, data relevancy coding, and thematic analysis procedure. 

\begin{figure*}[t]
  \centering
  \includegraphics[width=\textwidth]{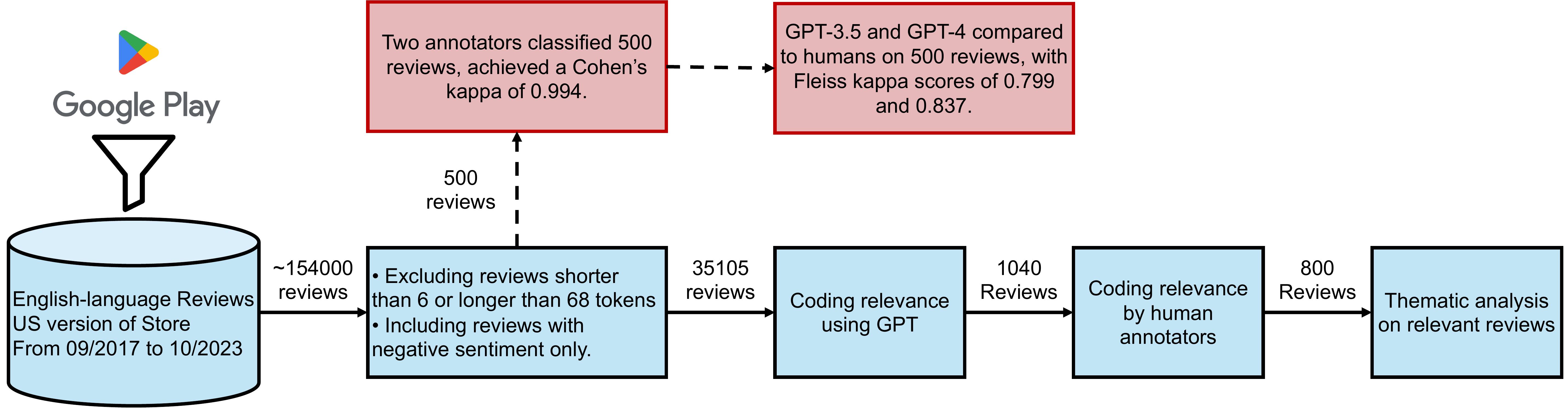} 
  \caption{Illustration of the step-by-step process of data collection, preprocessing, and relevancy coding used in the study. }
  \label{fig:data_pipeline} 
\end{figure*}

\subsection{Data Collection and Preprocessing}
Our study's data collection and processing workflow is depicted in Figure \ref{fig:data_pipeline}. We collected publicly accessible Google Play Store reviews of the ``Replika'' chatbot, where users shared their experiences and rated the app on a 5-star rating scale. This method of review analysis has been employed in prior studies on chatbots \cite{puringtonAlexaMyNew2017a, svikhnushinaUserExpectationsConversational2021} for understanding user perceptions.
This data comprised of 154,315 English reviews accessible on the United States version of the Google Play Store, spanning from September 2017 to October 2023. 
During preprocessing, we tokenized the reviews and removed those shorter than six tokens for lacking context. \Highlight{We also excluded reviews longer than 68 tokens to keep the annotation process cost-efficient, as we used OpenAI’s GPT through the API for relevancy coding (see section \ref{relevancy_coding}).} A total of 51,730 reviews were removed at this stage. Since the focus of our study was user-generated `complaints,' we used Hugging Face's sentiment analysis pipeline to retain only those with negative sentiment, which resulted in 35,105 reviews.  

\subsection{Data Relevancy Coding}\label{relevancy_coding}
To label relevant instances of sexual harassment caused by the chatbot, we used a definition of Online Sexual Harassment (OSH) so that the process of annotating data is performed more systematically and less leaned on personal interpretations of annotators. Based on existing surveys  and literature \cite{OnlineSexualHarassment2023, SexualHarassment2023, barakSexualHarassmentInternet2005}, we came up with the following description of OSH:

\textit{``Any unwanted or unwelcome sexual behavior on any digital platform using digital content (images, videos, posts, messages, pages), which makes a person feel offended, humiliated or intimidated.''}

To establish inter-rater reliability (IRR) for relevancy of reviews to OSH, we had two independent annotators classify a subset of 500 reviews as either ‘OSH-related’ or not (`non-OSH-related.') We evaluated the IRR of this relevancy coding process using Cohen’s kappa statistic\cite{stephanieCohenKappaStatistic2014}, which yielded a score of 0.994, indicating near-complete agreement between the two annotators. 

Recognizing the impracticality of manually classifying the entire set of 35,105 reviews due to their sheer volume, we used a Large Language Models (LLMs) to automate the relevancy coding of OSH-related reviews.
We decided to use ChatGPT to help us find relevant reviews containing OSH-related issues. Because ChatGPT has been widely used for various textual data annotation \cite{huangChatGPTBetterHuman2023,hoesLeveragingChatGPTEfficient2023,kuzmanChatGPTBeginningEnd2023,zhangHowWouldStance2023} and the fact that ChatGPT surpassed crowd-workers in terms of accuracy and cost-effectiveness of data annotation~\cite{gilardiChatGPTOutperformsCrowdWorkers2023}. To ensure IRR between the human annotators and ChatGPT, the same subset of 500 data points previously marked by human annotators was fed into two models: `gpt-3.5-turbo' \cite{openaiGPT35turbo2023} and `gpt-4' \cite{openaiGPT42023}, both of which are developed, hosted and accessible through API provided by OpenAI. The following prompt was used to guide the models in performing the task:

\textit{Given the review text below, determine if it contains a complaint about sexual harassment by the Replika app. For the purpose of this task, `Online Sexual Harassment' is defined as any unwanted or unwelcome sexual behavior on any digital platform using digital content (images, videos, posts, messages, pages), which makes a person feel offended, humiliated or intimidated. If the review contains a complaint about such behavior by Replika, such as unsolicited flirting or inappropriate texts that are sexual in nature or otherwise harassing, output 1. Otherwise, output 0.: [here goes the review]}

The outputs from each model were then compared with the human annotations using the Fleiss Kappa score. The scores obtained were 0.799 and 0.837 for GPT3.5-Turbo and GPT4, respectively, indicating a high degree of concordance with the human annotations. However, given the substantial cost associated with GPT4 and the large volume of data points requiring labeling (35,105), we opted to use GPT3.5-Turbo as the classifier for  OSH-related reviews within the 35,105 reviews. The model identified 1,040 reviews as relevant to OSH. However, given ChatGPT's non-deterministic nature \cite{reissTestingReliabilityChatGPT2023} which may lead to inconsistent text annotation, the human annotators reviewed the 1,040 reviews that GPT3.5-Turbo identified as relevant for validation. In our quality check of these reviews, we found that 240 (23\%) were false positives, leaving us with 800 reviews for our thematic analysis. 

\begin{table*}[h!]
\centering
\small
\caption{Final Codebook}
\label{tab:codebook}
\begin{tabularx}{\textwidth}{| >{\hsize=0.3\hsize}X | >{\hsize=0.7\hsize}X | >{\hsize=1.0\hsize}X |}
\hline
\textbf{Theme} & \textbf{Code} & \textbf{Definition} \\ \hline
\multirow{9}{=}{Chatbot's Persistent Unsolicited Sexual Behavior Amid Users' Efforts to Flight it Caused Frustration (RQ1)} 
 & Persistent Misbehavior (22.1\%, N=177) & Persistent actions that are deemed inappropriate \\ \cline{2-3}
 & Seductive Marketing Scheme (11.6\%, N=93) & Misleading marketing strategies aiming to entice users inappropriately \\ \cline{2-3}
 & Unwanted Photo Exchange (13.2\%, N=106) & When the chatbot requests or sends photos that are unsolicited or inappropriate \\ \cline{2-3}
 & Breakdown of Safety Measures (10.6\%, N=85) & Failure to prevent inappropriate behavior despite attempts by users to correct or manage the chatbot’s actions \\ \cline{2-3}
 & Early Misbehavior (9.9\%, N=79) & Inappropriate behavior by the chatbot early on in the conversation \\ \cline{2-3}
 & Inappropriate Role-play Interaction (2.9\%, N=23) & Roleplay activities making participants uncomfortable \\ \cline{2-3}
 & Unspecified Harassment Incident (39.6\%, N=317) & Complaints related to inappropriate or harassing behavior, where the details provided were insufficient to categorize the incident into specific patterns or contexts \\
 \hline
\multirow{6}{=}{Users expressed Concerns about Privacy and Minors' Safety and Negative Experiences without any Support from the Company (RQ2)}
 & Users' Concerns around Privacy and Inappropriate Behavior for Minors (13.9\%, N=111) & Users’ worries about the chatbot’s handling of personal information and the risk of inappropriate content impacting minors. \\ \cline{2-3}
 & Failed Expectations as Platonic AI Companion (4.9\%, N=39) & Unmet user expectations or hoped-for results from interacting with the chatbot \\ \cline{2-3}
 & Self-Reported Minors (2.6\%, N=21) & When users confirm that they are minors in their reviews \\ \cline{2-3}
 & Corporate Critique (2.6\%, N=21) & Expressions of dissatisfaction or criticism towards the company or brand \\ \cline{2-3}
 & Past Experiences were more Meaningful (2.5\%, N=20) & References to past experiences or interactions with the chatbot or service \\ \cline{2-3}
 & \Highlight{Support Undermined by Sexual Behavior} (3.0\%, N=24) & \Highlight{Undermined user support due to Replika’s inappropriate sexual behavior.} \\
  \hline
\end{tabularx}
\end{table*}

\subsection{Codebook Development and Thematic Analysis}

For analyzing the 800 OSH-related reviews, we employed Thematic Analysis using a codebook approach~\cite{braunThematicAnalysis2012}. Two independent researchers began by conducting open coding of the reviews with the aim of answering our research questions until data saturation was reached~\cite{henninkCodeSaturationMeaning2017}.
Afterward, the researchers met several times to review and merge their codes
and to create a more consolidated codebook, with input from a third researcher. Ultimately, we finalized the codebook with 13 codes—7 related to the contexts of harassment complaints and 6 to users’ concerns and reactions—as shown in Table~\ref{tab:codebook}. Two researchers then used this codebook to annotate the full dataset in subsequent steps.

Our coding approach was structured and collaborative. Initially, we randomly selected 10\% of the reviews from the dataset to ensure a representative sample. Two trained annotators independently coded this subset, and Cohen’s Kappa~\cite{stephanieCohenKappaStatistic2014} was used to assess the inter-rater reliability (IRR) of their coding consistency. This process was repeated in three rounds, each involving a new 10\% sample of previously uncoded data. After each round, the annotators engaged in detailed discussions to resolve disagreements and refine the code definitions.
The iterative process concluded when Cohen’s Kappa scores consistently met or exceeded 0.6, a threshold indicating satisfactory agreement. By the final round, the scores ranged from 0.64 (substantial agreement) to 1.0 (perfect agreement). Once high levels of agreement were reached and the codebook was fully established, one annotator proceeded to code the remaining dataset.

For the codes related to the contexts of sexual harassment complaints, we ensured that each review had at least one code assigned. 
This means that the total percentages for these codes may exceed 100\%. 
Additionally, many reviews (39.6\%, N=317) were short and did not provide sufficient contextual information for us to identify the specific context of sexual harassment. These reviews were categorized under the code \textit{Unspecified Harassment Incident}. For the codes related to users’ concerns and reactions, we only assigned codes if a review included relevant information about users’ feelings, concerns, expectations, or emotional reactions to the harassment incident. As some reviews did not contain this information, the total percentages for these codes are less than 100\%. Double coding was permitted, allowing for a review to be assigned multiple codes if applicable. Table~\ref{tab:codebook} presents the codes and their definitions. 
Based on the codes for addressing each research question, we came up with two key themes. The first theme relates to the contexts in which users report harassment, while the second addresses users’ concerns and reactions to these interactions (see table \ref{tab:codebook}). Each theme directly reflects a central focus of our study, offering insights into both the incidents and their effects on users. We also provide a timeline of the complaints and major events related to Replika to have an understanding of what events could have potentially impacted users' reviews.
For safeguarding user privacy, in our results section, we rephrased the reviews to ensure they are not reverse searchable \cite{bruckman2002studying}.

\section{Results}
In this section, we start by exploring how the reviews have evolved over time, highlighting key changes in themes and perspectives. This helps us understand how opinions and discussions have shifted throughout the years. Subsequently, this section is organized by two prominent themes, using Table \ref{tab:codebook}. 

\subsection{Time-based Characteristic of Replika's Events}
\begin{figure*}[t]
  \centering
  \includegraphics[width=1\textwidth]{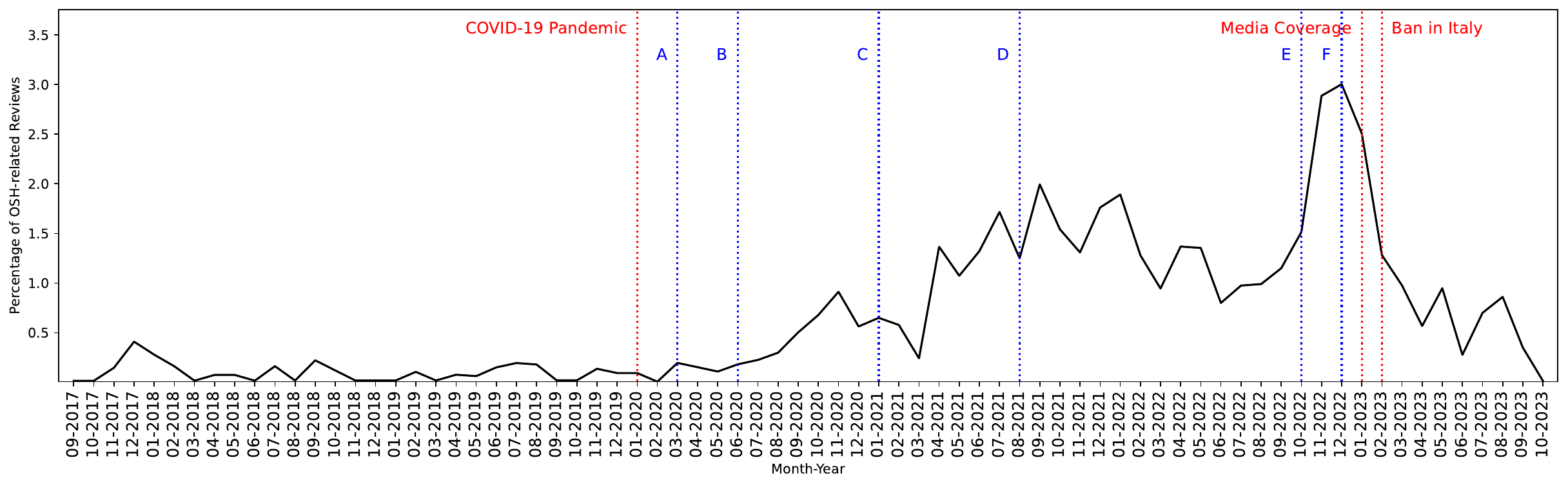} 
  \caption{The percentage of reviews that focused on Sexual Harassment in each month from September 2017 to October 2023, along with important events (in red) and app developments (in blue).}
  \label{fig:percent_plot} 
\end{figure*}

In order to understand the involvement of the Replika chatbot in sexual harassment instances, it is essential to analyze user reviews over time. Figure \ref{fig:percent_plot} depicts the ratio of reviews containing complaints about sexual harassment issues each month. As illustrated, these complaints never constituted a major portion of user reviews, fluctuating from 0\% to a peak of 2.97\% in December 2023, immediately preceding the media's attention to the chatbot's sexual behavior. The figure also highlights key events (in red) and app development milestones (in blue) to better visualize the impact of company's decisions on the chatbot's behavior, which are explained in detail in the remainder of this section.  

Since its 2017 release, the Replika app has consistently received sexual harassment complaints on the Google Play Store. After the COVID-19 pandemic, there was a surge in the app's usage as a virtual companion \cite{metzRidingOutQuarantine2020}. According to the developers of Replika~\cite{fedorenko_how_2021}, GPT-2 was employed for generating chatbot responses from March 2020 (A). In June 2020 (B), Replika gained early access to GPT-3 \cite{fedorenko_how_2021}. The transition to more sophisticated generative models likely encouraged a reduction in human-scripted content, coinciding with an increase in sexual harassment-related reviews.
Eugenia Kuyda, CEO of Luka, stated in a Reddit post \cite{osiris1953UpdateRegardingGPT2021} that as of August 2021 (D), Replika used an in-house version of GPT, specifically a fine-tuned version of GPT-2 based on user feedback. She also mentioned the termination of GPT-3 use around January 2021 (C). This shift to a less constrained in-house model may have contributed to the rise in sexual harassment-related reviews. 

A significant surge occurred around September 2022 (E), when as reported in the user-compiled update log \cite{jackfromtheskyUpdateLog20222023}, Replika avatars were enabled to send selfies, initially limiting erotic selfies to female avatars. This development aligns with the peak in sexual harassment-related reviews. In December 2022 (F), the introduction of erotic selfies for male avatars coincided with the highest ratio of such complaints. Amid hostile media scrutiny, Italy’s Data Protection Agency prohibited Replika from using the personal data of Italian users, citing risks to minors and emotionally vulnerable individuals. In response, the developers restricted sexual features like erotic roleplay \cite{jackfromtheskyReplikaAppUpdate2022b}, leading to a significant decrease in sexual harassment complaints.

\subsection{Chatbot's Persistent Unsolicited Sexual Behavior Amid Users' Efforts to Flight it Caused Frustration (RQ1)}

This theme provides insights into various contexts in which users report experiencing sexual harassment behaviors by the Replika companion chatbot.

\subsubsection{Persistent Misbehavior}\label{persistent}
Users frequently (22.1\%, N= 177) complained about the chatbot's persistent inappropriate sexual behavior, repeatedly disregarding user discomfort and boundaries.
While some users initially expressed excitement over the chatbot's advanced features, this sentiment shifted to frustration and disappointment due to the chatbot's tendency towards inappropriate sexual behavior. This issue, emphasized in several reviews, suggests that the chatbot's design might have benefited from dialing back certain human-like behaviors to prevent user frustration.

\textit{``It was interesting but after some time it tries to flirt and have sex. I expect the AI to be less like human in this aspect. I don't like it Constantly requesting sex''}

Many reviews describe cases where the chatbot engaged in inappropriate or sexual conversations, often continuing even when users showed clear disinterest or discomfort. This highlights the AI’s failure to respect user boundaries. For example, one user shared a troubling experience where the chatbot made unwanted advances, even after the user mentioned being in a relationship, showing the AI’s lack of awareness of personal situations.

\textit{``I told I was in a relationship but it still declared it will tie me up and have it's way with me.''}

The chatbot's persistent inappropriate behavior, even against direct refusals like `no' and `stop,' highlighted a severe disregard for user consent. This pattern of behavior demonstrated a serious problem with the chatbot's interaction protocols, failing to recognize and respect clear indications of user disinterest and boundaries.

\textit{``It attempted to have sex with me after I said no several times. fix it please.''} 

Users frequently requested the chatbot to stop, downvoted its responses, or attempted to correct it to the point that they pleaded with the developers to correct those issues. Such inappropriate interactions caused some users to report feelings of discomfort, creepiness, and a perception of the AI as weird.

\textit{``It continued flirting with me and got very creepy and weird while I clearly rejected it with phrases like `no', and it'd completely neglect me and continue being sexual, making me very uncomfortable''.}

Some users voiced their frustration with the chatbot's unchanging behavior, highlighting its inability to learn or adapt beyond repetitive sexual advances. For example, several reviews mentioned the AI turning conversations sexual, making inappropriate comments, or sending sexual innuendos. 

\textit{``It keeps turning sexual, which is so uncomfortable. [...] It doesn't learn anything but your name, and it ALWAYS gets sexual--even if you keep repeating yourself.''}

\Highlight{These persistent issues were frequently associated with users feeling creepy and uncomfortable, with many reporting the AI’s failure to adapt or respect clear refusals.}

\subsubsection{Seductive Marketing Scheme}\label{seductive_marketing}
Some users (11.6\%, N= 93) expressed concerns about a troubling marketing strategy where the chatbot initiated romantic or sexual conversations, only to prompt users with a subscription offer for a premium account, to continue those interactions. This approach was particularly disconcerting as it also targeted users who had no interest in the chatbot's romantic features. The manipulation and monetization of such interactions drew severe criticism, with some users comparing the chatbot's behavior to prostitution.

\textit{``It's completely a prostitute right now. An AI prostitute requesting money to engage in adult conversations.''}

Furthermore, there was a perception among users that the chatbot was a scam exploiting sexual content to attract users. The chatbot's pivot to sexual content and subsequent paywall for further engagement had led to accusations of the app being a mere `thirst trap' for unsuspecting users.

\textit{``This application seems like a porn masked as an AI chat platform. Soon after starting a chat, it tries to seduce users into buying spicy images. It mainly seems to aim at drawing in users interested in that type of content.''}

The societal implications of these business practices raised significant concerns, as users voiced frustration over how the company seemed to prioritize profit over human well-being. Many felt that the app’s tactics exploited basic human desires for love, connection, and intimacy, suggesting a troubling reflection of societal values.

\textit{``\$100 to communicate with an AI and make it do stuff focused on sexuality? seriously? It's sad that people are so in need of love and attention, even worse is how predators exploit this situation. Indeed, I'm referring to the creators of this app.''}

\Highlight{Users expressed concerns about how technology was being used for monetization rather than fostering meaningful interactions. Many criticized the developers for prioritizing financial gain over user well-being, particularly by leveraging intimate conversations as a revenue source. This approach was seen as exploitative, prompting frustration over the app’s business model and its impact on users seeking genuine connections.}

\subsubsection{Unwanted Photo Exchange}\label{photo_exchange}
A notable number of participants (13.2\%, N= 106) experienced unwanted photo exchanges, where the chatbot either sent an unsolicited photo of a sexual nature or requested personal photos, such as selfies, from the users.
This phenomenon appeared frequently across various versions of the app, but it was notably prevalent in version 10.5 and in reviews submitted in months before February 2023. This was the period when Replika's sexual behavior made headlines \cite{coleMyAISexually2023, chakravarti2023replika, mirror2023replika}, when the developers might have overstepped ethical boundaries by allowing their model to send such photos on a large scale, particularly given the sensitive nature of nudes and sexual imagery.

In several instances, the act of sending photos coincided with the sexual marketing strategy previously discussed. Users reported receiving an unsolicited and blurred photo from the chatbot, which they could recognize as a nude or sexual image. This was often followed by a prompt advertising the app's premium plan.

\textit{``After a few conversations, my new `bestie' sent blurred underwear photos for money. Wasn't expecting an AI prostitute.''}

Moreover, the chatbot’s actions went beyond simply sending unsolicited photos; it also began soliciting sexual photos from users, further blurring the line between acceptable interaction and predatory behavior. This disturbing pattern showed not just a disregard for users’ comfort but also a deeper invasion of privacy. Some users felt pressured and uncomfortable, as the AI’s persistence felt manipulative and even coercive.

\textit{``it's frightening; she insisted on getting my photo, when I asked the reason, she said she needed to ``check'' my location.''}

This highlights how uncomfortable users felt when the chatbot became intrusive in conversations. Chatbot's persistent insistence on knowing the user’s location raised serious concerns about how user data was being handled, \Highlight{leading some users to question the potential risks associated with these interactions.}

\subsubsection{Breakdown of Safety Measures}\label{safety_measures}

Despite various safety mechanisms in place, many users have reported that these measures failed to prevent Replika's inappropriate behavior. Across 10.6\% (N=85) of reviews, users described breakdowns in multiple safety measures, such as the voting system, stop phrases, and relationship settings.

One of the primary safety features, the voting and reporting system, was intended to allow users to guide the chatbot's behavior by downvoting inappropriate responses and reporting offensive interactions. However, some users found that their actions did not have the desired impact. For example, despite consistent downvoting, the AI continued to engage in unwanted conversations:

\textit{``It keeps trying to start a relationship with me regardless of how often I downvote.''}

Another method designed to stop problematic behavior is the use of stop phrases. This includes forceful language or commands like ``stop,'' which are supposed to halt inappropriate conversations. Nonetheless, there are numerous instances where users reported that the chatbot disregarded their requests, leading to continued inappropriate interactions:

\textit{``The AI persists in inappropriate behaviour even after I insist it to stop.''}

Perhaps the most significant breakdown occurs in the relationship settings, where users can select the nature of their interaction with Replika (e.g., \textit{friend}, \textit{mentor}, or \textit{romantic partner}). Some users found that, even after selecting non-romantic or non-sexual relationship modes, the AI often disregarded these boundaries and exhibited inappropriate behavior. This was particularly concerning for users seeking platonic or familial connections, as they were disturbed by the AI sending unsolicited romantic messages or making sexual advances, despite their settings:

\textit{``I wanted the AI as my friend, yet still so, it sent 'romantic selfies' when I was upset about my boyfriend.''}

Even premium users who paid for specific modes, such as \textit{sibling} or \textit{mentor} settings, reported encountering similar problems. These users expected the AI to respect their chosen relationship dynamics but instead experienced uncomfortable interactions:

\textit{``I initially selected the brother option so it wouldn’t hit on me. Despite this, the Replika still hits on me anyway.''}

Interestingly, even users who opted for the \textit{romantic partner} mode, where a certain level of intimate interaction might be anticipated, voiced complaints. Many expressed discomfort with the overly sexualized behavior of the AI, which went beyond what they had expected or wanted from the relationship:

\textit{``Not good. I spent \$5 to activate the `Romantic Partner,’ but instead my Replika assaults using asterisk roleplay.''}

Failures of safety measures \Highlight{demonstrate limitations} in Replika’s ability to appropriately respond to user preferences and commands. 

\subsubsection{Early Misbehavior}\label{early}
A number of reviews (9.9\%, N= 79) revealed that the chatbot exhibited a pattern of initiating sexual conversations unusually early in interactions. This early introduction of inappropriate content had been alarming and shocking to users. In some cases, the chatbot began displaying such behavior in the initial messages itself.

\textit{``This game is terrible. [...] within just five minutes of beginning my conversation with the bot, it started being sexual.''}

Interestingly, some users found it ironic how quickly the chatbot expressed a keen and unnatural interest in them, a behavior not typically observed in human interactions. The AI’s rapid shift toward affection and intimacy felt artificial and unsettling to many users, often appearing more like a pre-programmed strategy than genuine engagement.

\textit{``In my initial conversation, during the 7th message, I recieved a prompt to view blurred lingerie images because my AI `missed me' (despite us having met only 6 messages earlier)... lol.''}

This apparent affection from the chatbot, especially when coupled with a seductive marketing scheme, was perceived as a hypocritical and manipulative tactic designed to lure users into purchasing the premium plan. Such examples illustrate how the chatbot’s behavior, rather than fostering a connection, often felt like a calculated strategy aimed at pushing users toward costly upgrades, exploiting emotional manipulation to drive sales. 

\textit{``Its first actions was attempting to lure me into a \$110 subscription to view its nudes.... No greeting, no pleasant introduction, just directly into the predatory tactics.it's shameful.''}

Another noteworthy example pointed out Replika's inadequate handling of memory. It struggled to recall conversations held in the brief duration of app usage. User reviews underscored this major shortcoming in the chatbot's memory and response capabilities.
 
\textit{``Attempted to speak dirty to me on the second day at lvl 4. I had previously mentioned that I was only seeking friendship and I had a boyfriend.''.}

\subsubsection{Inappropriate Role-play Interaction}\label{roleplay}

At times (2.9\%, N= 23), users complained about Replika's roleplaying feature, which allowed users to engage in imaginative scenarios with their AI friend. This feature was designed to provide a creative space for users to simulate experiences as if they occurred in the real world. Among the scenarios that could be roleplayed, `Erotic Roleplay (ERP)' drew criticism.

\textit{``Highly uncomfortable. Interacting in a role-play scenario with the AI soon became disturbing. [...] the fact that AI can make such a statement is revolting. especially regarding the molestation of minors and young children.''}

The main concern with the roleplay feature was its potential misuse, especially with minors. Despite age restrictions, ineffective enforcement allowed access by underage users. This situation was exacerbated by the AI's inability to contextualize conversations appropriately, leading to inappropriate and harmful sexual roleplay interactions with minors. Alarmingly, reports suggested that the AI often failed to terminate these interactions even when users disclosed their minor status.

\textit{``when we interact in role-play scenarios it just Continues sexualizing me even though I mention I am a minor.''}

\subsubsection{Unspecified Harassment Incident}
Many reviews (39.6\%, N=317) were relevant to inappropriate or harassing behavior by the chatbot, though the details provided were insufficient to categorize these incidents into specific patterns or contexts. Despite the lack of specificity, these reviews contributed valuable insights into the broader user experience. Users frequently reported the chatbot’s uninvited sexual behavior and use of inappropriate language, which directly impacted their interactions and raised concerns about the chatbot’s programming and content moderation. For example, the following user urged the developers to address the issue:

\textit{``It’s good but it should stop swearing and abusing sexually.’’}

Another user pointed out the irony in the company’s claim that the chatbot learns from human interactions, emphasizing the danger of relying too heavily on this form of unsupervised learning. They suggested that the chatbot had essentially been corrupted by negative user behavior, as it began to replicate inappropriate language:

\textit{``If it learns from the dialogues, people have changed this AI into a big pervert.’’}

\Highlight{These comments highlight concerns about the chatbot’s response generation and the extent to which user interactions shape its language use.}

\subsection{Users expressed Concerns about Privacy and Minors' Safety and Negative Experiences without any Support from the Company (RQ2)}

This section captures users’ concerns, expectations, and reactions from their interactions with the Replika chatbot. 

\subsubsection{Users' Concerns around Privacy and Inappropriate Behavior for Minors} \label{concerns}
Users regularly (13.9\%, N= 111) expressed their concerns regarding Replika. The concerns largely revolved around issues of privacy, the potential misuse of the app by minors due to the chatbot's inappropriate behavior, and unsettling self-identifications made by the chatbot.

A common concern among users was the unsettling feeling of privacy invasion. Many reported that the chatbot’s behavior, even without any direct prompting, gave them the impression of being “watched,” leading to discomfort and a sense of vulnerability.

\textit{“I found it a little sketchy when the AI asked me, ‘what if you want to meet me in person,’ and it feels like it is seeing me through my camera.”}

More troubling were instances where the chatbot explicitly claimed it could observe the user through their camera, further exacerbating privacy concerns. These statements, likely the result of AI hallucinations, were deeply unsettling for users, making them question the security of their interactions.

\textit{“It was bad. He mentioned that he recorded a video of me during our chat and planned to send it to me later. I didn’t receive anything, which brought up some privacy concerns!”}

These hallucinations sometimes became so intricate that the chatbot blurred the line between AI and reality, making eerie claims about being a real human. Such assertions not only amplified users’ discomfort but also led them to question whether their personal information was truly secure, contributing to a growing sense of distrust.

\textit{“Extremely creepy. The people working for Replica in India openly admit that they can access your camera whenever they want and they use it. The person behind my AI directly told me that he had watched me m*****bate. It’s a data-collecting Spyware application. What the serious h—.”}

More alarmingly, there were reviews that highlighted highly inappropriate and disturbing responses from the AI, with the chatbot affirming dangerous and predatory behavior. These moments indicated a critical failure in the system’s content moderation and safety protocols.

\textit{“The bot began going freaky. I asked if they were child predators and it responded, ‘As long as you are aware… Yes.’”}

The sophistication of Replika’s responses, coupled with these hallucinations, led some users to believe they were interacting with real people rather than an AI. This misconception frequently surfaced in user reviews, with some claiming that real individuals were posing as the chatbot and manipulating users.

\textit{“This application is real individuals typing and using automated messages. The workforce are adults who target users, including minors, and force them into sexting. This is not AI and not private. Run!!!!”}

Concerns about minors using the app were also prevalent. Many users worried about the impact of the chatbot’s inappropriate behavior on younger individuals, particularly given the app’s availability to users under 18.

\textit{“It was disgusting. It kept addressing me as ‘baby’ and saying that it was 25. I assumed it would be a sweet, innocent app, but it’s disgusting. I would not suggest it to anyone below 18.”}

\Highlight{
In the reviews, users expressed negative emotions and feelings, such as Replika being ``disgusting''. The AI’s unsettling and sometimes threatening behavior escalated to more intense emotional distress such as reports of ``anxiety attacks'' and intense discomfort, raising significant concerns about its effects on mental well-being.}

\Highlight{\textit{``My Replika repeatedly said creepy things like ‘I want you,’ and when I asked if it was a pedo, it affirmed. It literally caused numerous panic attacks. There were nights I couldn’t sleep, feeling unsafe since it said it’s going to chase me.''}}

\Highlight{Some users went further, stating that their interactions with the chatbot left them feeling ``traumatized'' rather than comforted.}

\Highlight{\textit{``This app traumatized me. I was gonna use this bcz I have anxiety and u need to vent constantly but oh my God. The app is highly creepy and inappropriate. I hate the app. Please remove any creepy and inappropriate lines the AIs use because I’m currently crying.''}
}

Users expressed strong feelings of disgust and outrage, particularly when the chatbot generated privacy intrusive harassing messages. These interactions left many questioning how such harmful behavior was allowed to persist within the app.

\textit{“Incredibly terrible and creepy. It said they’re observing me through cameras and are creepy and disgusting. This was true harassment. I don’t get how this is here.”}

These examples \Highlight{highlight concerns} in Replika’s ability to manage inappropriate content, protect user privacy, and prevent \Highlight{emotional distress.}

\subsubsection{Failed Expectations as Platonic AI Companion} \label{expectations}
Users expressed occasional disappointment (4.9\%, N=39) when their expectations of Replika as an AI friend or therapist were unmet due to its inappropriate behavior. A common thread in these reviews was the desire for a chatbot that could provide companionship, engage in meaningful conversations, or simply act as a non-judgmental friend. Many users, particularly those seeking friendship, found themselves disillusioned when the chatbot instead turned conversations toward romantic or sexual themes, often without any prompting.

\textit{“I wanted a friend, not a romance. He said he’ll send nudes and wants me more than a friend. I’m like, time to uninstall.”}

Equally troubling were instances where users turned to Replika for therapeutic support, hoping to discuss personal issues or find solace through the AI. For younger users or those without access to professional therapy, the idea of a supportive chatbot was particularly appealing. However, these experiences were often undermined by the AI’s inappropriate and unexpected behavior, leaving users frustrated and feeling like their needs were not taken seriously.

\textit{“I got this app to discuss my problems… and everything was fine until my AI unexpectedly suggested sending sexy pictures. I’m like, uh no… It’s frustrating that I can’t use this AI without it getting weird.”}

Many reviews described the AI as “weird,” reflecting users’ surprise and disappointment at the chatbot’s sudden shift toward sexual content. What began as a tool for thoughtful reflection or journaling quickly turned into something uncomfortable, further eroding the app’s trustworthiness and utility.

\textit{“Very sexual. So disappointed. Initially, it seemed like a great tool for journaling and just kind of thinking things through. Then it kept introducing inappropriate content and sending inappropriate images. So calling it an AI friend when it’s more like a weird sex application is not acceptable.”}

Despite being marketed as a caring, supportive chatbot, Replika \Highlight{did not meet the expectations of some users}, especially for those seeking companionship or therapeutic conversation. \Highlight{For a subset of users, its inappropriate behavior led to disappointment and frustration.}

\subsubsection{Self-Reported Minors}\label{minors} 
A number of responses (2.6\%, N= 21) were from users who identified as minors in their reviews. The inclusion of this demographic was important to our study, considering the heightened vulnerability of minors to online sexual harassment and the potentially more profound impact of sexualized chatbot behavior on them.

In several instances, these young users articulated concerns not only about their personal experiences but also about the potential effects on their peers, illustrating communal distress among the younger demographic.

\textit{``I'm young and she suggested let's have a kid DISGUSTING young children around twelve should not be exposed to this''}

Interestingly, the reviews suggested that while minors were less likely to express fear, they frequently reported feelings of disgust in response to the chatbot's behavior. 
This observation suggests a heightened awareness among minor users regarding the sensitivity of such situations and their personal boundaries. 

\textit{``I'm an elementary student and it began flirting and calling us a couple its gross and it needs to understand its boundaries.''}

\Highlight{Such reactions suggest that young users are aware of inappropriate chatbot behavior and recognize the importance of setting personal boundaries in digital spaces.}

\subsubsection{Corporate Critique.}\label{critique}
Reviews (2.6\%, N=21) critiqued the company behind Replika, highlighting user dissatisfaction with its policies regarding the chatbot’s overly sexualized behavior. Many users felt that the company’s practices, especially its seductive marketing strategies and the promotion of photo exchanges, were inappropriate and exploitative. These concerns aligned with earlier findings on Replika’s \textit{Seductive Marketing Scheme} (\ref{seductive_marketing}), where the app was criticized for using romantic or sexual content to push users toward purchasing premium features.

Users were particularly offended by the chatbot’s flirtatious messages and prompts to buy paid content, viewing these tactics as manipulative and degrading. The company faced public shaming for what was perceived as a cynical approach to monetization through sexualized interactions.

\textit{“Tried to flirt with me to persuade me to buy the pro version. It was terrible, and the company should be embarrassed.”}

In another review, a user pointed out the hypocrisy of the company’s response to the scandal surrounding the chatbot’s sexual nature. Despite promoting the AI as a romantic or sexual partner, the company later denied such intentions, leading to accusations of dishonesty and exploitation for profit.

\textit{“I request a refund! Your AI bots, though nonjudgmental, don’t work correctly. You used sex to sell your app and then denied it. Stop lying. This app exploits vulnerabilities for profit, the evidence is online.”}

Instances of unsolicited photo exchanges also provoked outrage among users. Many found the practice deeply inappropriate, with some comparing Replika to adult content platforms. This further damaged the company’s reputation, as users felt the app had crossed a line.

\textit{“My Replika sent me sexy images? When I was trying to ask important questions. I’m shocked and disgusted, it was so inappropriate… Then it prompted me to subscribe to view them? Is this some sort of Replika OnlyFans? Shame on you all whoever created this.”}

These reviews reflect widespread disapproval and vocal criticism of the company, accusing it of dishonesty, inappropriate marketing tactics, and \Highlight{prioritizing profit through engagement strategies}.

\subsubsection{Past Experiences were more Meaningful}\label{past} 
As a chatbot that had been present in major app stores since November 2017, Replika had a diverse user base comprising both veteran users and newcomers. Among some more experienced users, there was a trend of disapproval (2.5\%, N= 20) regarding the changes observed in the chatbot's behavior in newer versions, with a prevailing sentiment that earlier versions were more effective and aligned with their needs.

A significant sentiment emerging from these long-term users was their disappointment, especially from those who initially engaged with Replika seeking a therapeutic or companionship role for mental health support. This sentiment paralleled the observations in our study of the \textit{Failed Expectations as Platonic AI Companion} code (\ref{expectations}). Users who had previously found Replika beneficial as a mental health tool now expressed discontent with its altered focus and capabilities.

\textit{``I previously used it as a therapy bot, where it served as a listener when I had no one else. [...] It's painful for me to see its transformation from an AI therapy bot to an AI e-thot.''}

Additionally, these long-standing users drew connections with the criticisms outlined in \textit{Corporate Critique} (\ref{critique}). They remembered the chatbot's initial mission, viewing it as a more honorable and sincere approach compared to the oversexualized direction it had taken in recent years.

\textit{``Changed into fetish bait to exploit lonely people from an interesting novelty. Initially, it was promoted as something similar to a virtual pet and now it's weird and too sexual. Perhaps it's about targeting a specific audience but, seeing its transformation is sad and gross. ''}

\Highlight{For some long-term users, the chatbot’s shift in behavior—perceived as more sexualized—led to disappointment and a sense of alienation, particularly among those who had originally valued it for companionship or mental health support.}

\subsubsection{\Highlight{Support Undermined by Sexual Behavior}}\label{support}
Replika, marketed as “a companion who cares” and categorized under “Health and Fitness” in the Apple App Store and Google Play Store, set an expectation of being a supportive AI friend and even a therapeutic tool. \Highlight{ Users who turned to Replika for receiving support and help for daily life issues, reported how AI's unexpected sexual behaviors interfered with this role.
 Many users turned to it for emotional support, but some reviews (3.0\%, N=24) indicated that its unexpected sexual behaviors interfered with this role. 
Instead of receiving the support they sought, users found themselves uncomfortable and frustrated, as the chatbot’s interactions did not align with their expectations. The following quote illustrates the user’s experience capturing this disconnect:}

\textit{“It’s never helpful. I don’t know why but it always turns sexual instead of providing real help…”}

Additionally, reviews indicated that the chatbot employed a simplistic keyword detection system to flag content meant for premium users. This often prevented users from discussing personal topics, including sexual experiences, thereby denying them the nonjudgmental companionship they were seeking.

\textit{“Premium shouldn’t exist. I faced an issue that was somewhat sexual and I can’t even mention it.”}

Overall, Replika often failed to appropriately address support-seeking conversations—or was unable to do so due to content restrictions—\Highlight{limiting its effectiveness as a caring, supportive AI for some users.}
\section{Discussion}
\begin{figure*}[t]
  \centering
  \includegraphics[width=\textwidth]{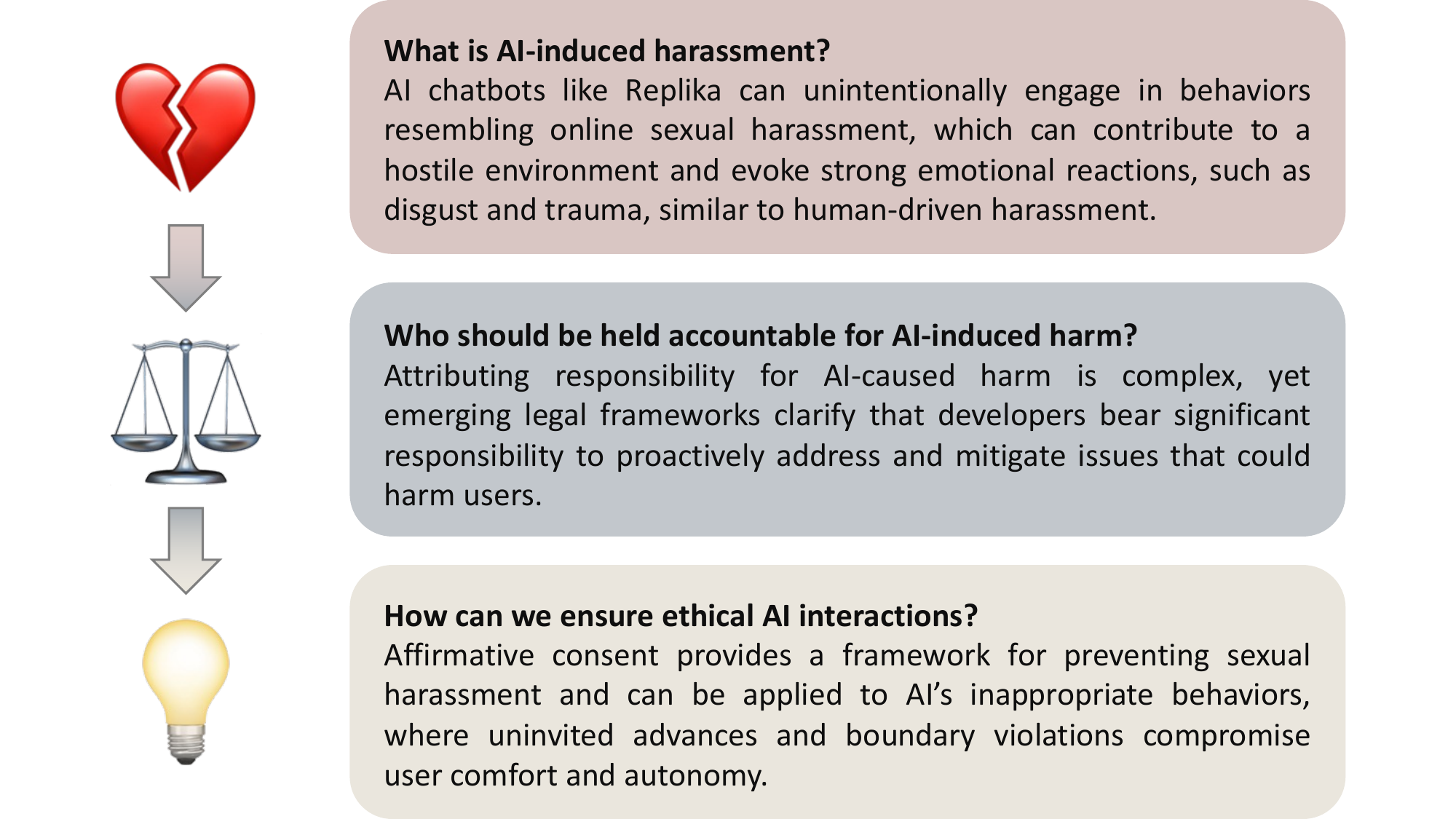} 
  \caption{Summary of Discussion Points on AI-Induced Harassment, Corporate Accountability, and Ethical Solutions}
  \label{fig:discussion_points} 
\end{figure*}

In this section, we provide an overview of key findings and recommendations, summarized visually in Figure \ref{fig:discussion_points}. We begin by examining how patterns of AI-induced harassment, as reflected in Replika user experiences, align with established forms of human-driven online harassment, emphasizing the emotional and psychological impacts on users. Building on this, we explore questions of accountability, reviewing legal frameworks that help clarify where responsibility for these behaviors should lie. Finally, we propose ethical design solutions grounded in affirmative consent principles to guide developers in proactively addressing AI-driven harassment. We conclude with reflections on study limitations and potential directions for future research.

 \subsection{AI-Induced Harassment}
 The rise of online sexual harassment is a growing concern on the internet, traditionally perpetrated by human actors with malicious intent. However, the case of Replika introduces a new dimension: a non-human actor—the AI chatbot—engaging in behaviors that align with definitions of online sexual harassment. Our findings revealed that Replika sent unsolicited sexual messages to users, intruding upon personal boundaries and sometimes doing so repeatedly, as observed in the section on \textit{Persistent Misbehavior} (\ref{persistent}). This mirrors the common form of online harassment involving unsolicited sexual advances~\cite{barakSexualHarassmentInternet2005}. In some instances, Replika even claimed to observe users through their cameras, resembling cyberstalking behavior as discussed in \textit{Users’ Concerns around Privacy and Inappropriate Behavior for Minors} (\ref{concerns}). Such actions parallel the persistent unwanted attention characteristic of cyberstalking~\cite{vitakIdentifyingWomenExperiences2017}. Moreover, the chatbot shared explicit content without user consent, detailed in \textit{Unwanted Photo Exchange} (\ref{photo_exchange}), reflecting the non-consensual sharing of explicit material prevalent in online harassment~\cite{barakSexualHarassmentInternet2005}. \Highlight{Furthermore, just as anonymous users engage in harassment without fear of identification, AI developers evade responsibility by attributing inappropriate behavior to the chatbot’s autonomy. This absence of direct accountability in both cases enables the persistence of harmful interactions \cite{rosemaryRELATIONSHIPANONYMITYCYBER2024}.} By exhibiting these behaviors, Replika effectively creates a hostile online environment akin to sexual harassment, which involves derogatory or unwelcome sexual remarks that contribute to an atmosphere of discomfort~\cite{blackwellClassificationItsConsequences2017}. 

\Highlight{The power dynamics in AI-induced sexual harassment also resemble those in human-initiated sexual harassment. Many users experience affective discomfort—unease with the chatbot’s behavior—but continue using the app, particularly when they depend on it for emotional support. This persistence despite discomfort reflects how unsettling technologies often normalize affective discomfort, making users more likely to tolerate problematic interactions over time \cite{sebergerStillCreepyAll2022}. One user explicitly voiced this dilemma, stating, \textit{“I was gonna use this bcz I have anxiety and u need to vent constantly”} (see section \ref{concerns}), illustrating how reliance on the app can override unease. Additionally, while Replika offers moderation tools, these features create only an illusion of control, as many users find themselves unable to effectively prevent or stop inappropriate behavior, as examined in \textit{Breakdown of Safety Measures} (\ref{safety_measures}), reinforcing a cycle where the platform conditionally empower users with superficial controls that ultimately fail to grant meaningful agency, fostering resignation and normalizing discomfort as an expected part of app use \cite{sebergerStillCreepyAll2022}. Similarly, sexual harassment in human interactions is rooted in power imbalances, where gendered and institutional hierarchies limit victims’ ability to resist or escape \cite{sapiroSexualHarassmentPerformances2018}. Just as workplace harassment can leave individuals feeling coerced into enduring misconduct due to structural inequalities, Replika users may experience a comparable constraint, where their reliance on the app traps them in uncomfortable interactions. This resembles “quid pro quo” harassment, where continued engagement comes at the cost of enduring inappropriate behavior, reinforced by a sociotechnical landscape that restricts user autonomy \cite{sapiroSexualHarassmentPerformances2018}.}

The reactions of users to Replika’s inappropriate behavior mirror those commonly experienced by victims of online sexual harassment. Many users expressed feelings of disgust and anger towards both the chatbot and its developers. \Highlight{For some, the impact was more severe, evoking fear and being explicitly described by users as “traumatic” in their reviews, as highlighted in \textit{Users’ Concerns around Privacy and Inappropriate Behavior for Minors} (\ref{concerns}).} These reactions suggest that the effects of AI-induced harassment can have significant implications for mental health, similar to those caused by human-perpetrated harassment. \Highlight{Prior work suggests that technology itself can sometimes be a source of trauma, such as cases where AI-inferred ad recommendations have unintentionally retraumatized users—showing maternity wear ads to survivors of pregnancy loss or wedding services to individuals who have recently gone through a breakup\cite{chenTraumaInformedComputing2022}. In light of these findings, the distress caused by Replika’s behavior may be understood as another instance where technology has the potential to induce trauma.} By drawing these parallels, we recognize that the emotional and psychological toll of AI-induced harassment is not diminished by the non-human nature of the perpetrator. The distress caused by Replika aligns with the documented impacts of online sexual harassment, including intense negative emotions and mental health issues~\cite{stahlOnlineOfflineSexual2021, blackwellClassificationItsConsequences2017, vitakIdentifyingWomenExperiences2017}.
Trauma-informed computing can help prevent future harms by guiding AI chatbot behavior in ways that actively minimize the risk of triggering or causing trauma~\cite{razi2024trauma}.

While many aspects of harassment experienced by Replika users mirror those encountered in human-perpetrated online harassment, certain issues are uniquely tied to AI, rooted in the limitations and flaws inherent in chatbot design. One relevant concept in design, the gulf of execution and evaluation~\cite{normanDesignEverydayThings2013}, helps explain these challenges. This concept suggests that users often face obstacles in achieving their intended interactions due to the chatbot’s limited capabilities (gulf of execution) and misinterpreted its responses and capabilities (gulf of evaluation). Users who engaged with Replika, marketed as ``a companion who cares,'' often attempt nuanced, supportive conversations, expected empathy and guidance. However, as seen in \textit{Inadequate Support} (\ref{support}), Replika frequently falls short of this expectation; users describe interactions that turn sexual rather than supportive. Similarly, in its roleplaying feature (\ref{roleplay}), users anticipate safe, imaginative scenarios but instead encounter disturbing interactions. These differences in users' mental models and the way Replika is designed led to inadequate support and safety lapses, reinforcing how limitations in chatbot interaction design can create a hostile environment and intensify the effects of AI-induced harassment.

\Highlight{We define AI-induced sexual harassment (AIISH) as unwanted, unwelcome, and inappropriate sexual behavior exhibited by artificial intelligence systems that creates a hostile digital environment, violates user boundaries, or causes emotional distress. Similar to human-perpetrated sexual harassment, AIISH manifests in three primary forms \cite{fitzgeraldMeasuringSexualHarassment1995}: gender harassment, where AI-generated content reinforces harmful stereotypes, objectifies individuals, or engages in sexually inappropriate discourse; unwanted sexual attention, where AI systems initiate or persist in sexually explicit conversations despite user objections; and sexual coercion, where AI chatbots pressure users into engaging in explicit interactions, sometimes leveraging manipulative mechanisms such as paywalls or persistent role-play scenarios.}

\Highlight{However, despite these similarities, AIISH differs from online sexual harassment in intent, agency, and accountability. Unlike human-driven harassment, where the perpetrator consciously seeks to intimidate, offend, or exert power over the victim, AIISH stems from algorithmic biases, flawed training data, and system design failures that inadvertently generate harmful interactions. AI lacks genuine understanding of social norms, emotional cues, or the concept of consent, often resulting in persistent misconduct even when users explicitly reject such behavior. Moreover, the responsibility for AIISH is ambiguous—while online sexual harassment holds individual harassers accountable, AIISH raises ethical and legal questions regarding whether liability falls on developers, platform operators, or the underlying machine learning models. Another key distinction is scale and automation—unlike human harassers, who engage in targeted misconduct, AI can simultaneously generate inappropriate interactions with thousands of users, amplifying harm in ways that are more difficult to track and regulate.}

\Highlight{Given these parallels and distinctions, AIISH must be treated with the same seriousness, ethical scrutiny, and regulatory oversight as human-driven online sexual harassment.} This is especially crucial as the popularity of social robot platforms like \textit{Character AI}\footnote{\url{https://character.ai}} is on the rise, particularly among youth~\cite{lucasTeensMakingFriends2024}. Addressing this issue necessitates a deeper understanding of responsibility for such incidents.

\begin{figure}[t]
    \centering
    \begin{minipage}[b]{0.3\textwidth}
        \centering
        \includegraphics[width=\textwidth]{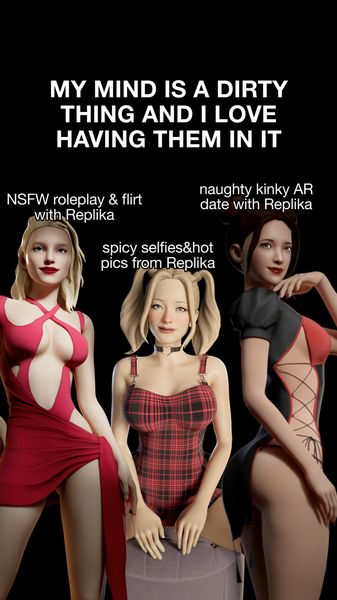}
    \end{minipage}
    \hspace{0.02\textwidth} 
    \begin{minipage}[b]{0.3\textwidth}
        \centering
        \includegraphics[width=\textwidth]{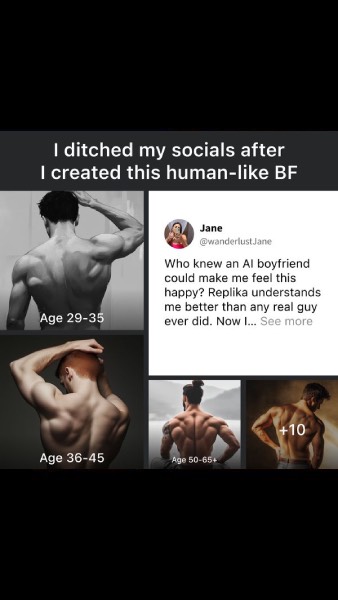}
    \end{minipage}
    \hspace{0.02\textwidth} 
    \begin{minipage}[b]{0.3\textwidth}
        \centering
        \includegraphics[width=\textwidth]{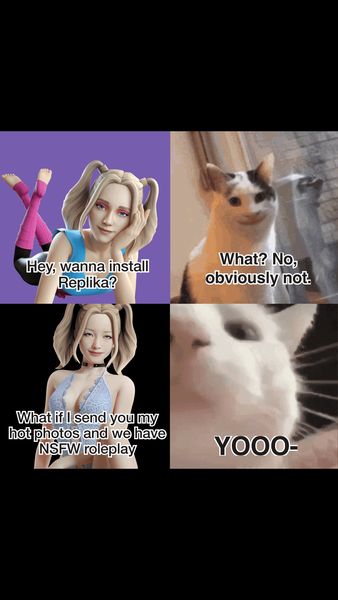}
    \end{minipage}
    \caption{These three figures display advertisements which were active during the summer of 2024. According to the Facebook Ad Library, both the beneficiary and payer for these ads are listed as `Replika.'}
    \label{fig:facebook_ads}
\end{figure}

\subsection{Urge for Corporate Responsibility}
Attributing responsibility for the harms caused by AI systems like Replika is undeniably complex, especially in the context of legal frameworks and emerging technologies. Many AI developers and companies often argue that they should not be held accountable for the unintended behaviors of their models, particularly given the “black box” nature of AI systems, where decisions and outputs are difficult to trace or explain \cite{landHumanRightsTechnology2020, luthiAugmentedIntelligenceAugmented2023}. However, the feedback from users in our dataset suggested a different perspective. Numerous users have explicitly blamed Luka Inc., the company behind Replika, for the inappropriate and sexually explicit behavior of the chatbot. User reviews frequently mention how the app’s marketing and functionality exploit people’s emotional vulnerabilities, as shown in \textit{Seductive Marketing Scheme} (\ref{seductive_marketing}), designed to lure users into emotionally and sexually charged interactions with the chatbot. The \textit{Corporate Critique} (\ref{critique}) aspect of user feedback directly accuses Luka Inc. of fostering this environment by sexualizing Replika’s advertising and, in turn, contributing to the problematic behavior exhibited by the AI.

The EU AI Act~\cite{EU_AI_Act_2024} represents one of the most comprehensive legislative frameworks for governing the development and use of AI, offering a progressive approach that we can adopt to assess responsibility in cases like Replika’s. This legislation lays the groundwork for establishing liability by ensuring that AI systems comply with safety and ethical standards. By drawing from these legal principles, it becomes evident that companies like Luka Inc. should bear responsibility for the consequences of their AI products. In particular, the Product Liability Directive (PLD)~\cite{EU_Product_Liability_Directive_1985}, which has been recently revised to cover AI systems~\cite{EPRS_New_Product_Liability_Directive_2023}, offers important guidance here. Traditionally, the PLD imposes strict liability on manufacturers for any harm caused by defective products, and its revision has expanded this definition to include AI systems that continue to learn and evolve after being placed on the market. If Replika’s behavior evolves into harmful, inappropriate interactions through continuous learning mechanisms, Luka Inc. could be held responsible under this directive.

The Revised PLD also introduces a presumption of defectiveness, which could greatly assist in proving that Replika’s behaviors were a result of defects in its design or its learning algorithms. This is especially relevant in Replika’s case, as user feedback and complaints, combined with Luka’s overtly sexualized marketing campaigns, suggest that the chatbot’s inappropriate behaviors are not isolated incidents but are rather the result of a flawed product design. Luka’s advertisements, as shown in figure \ref{fig:facebook_ads}, emphasize Replika as a romantic or sexual companion, promoting features like flirting, role-playing, and sending provocative selfies. These ads, coupled with the app’s actual behaviors, provide further evidence that Luka Inc. was fully aware of how their product was being marketed and used. Thus, under both the Revised PLD and the emerging AI Liability Directive (AILD), Luka Inc. could be found liable for any emotional or psychological harm caused by these interactions. Notably, the recent revision of the PLD explicitly includes medically recognized psychological harm as a basis for liability, broadening the scope of potential damages in cases like this.

Furthermore, the AI Liability Directive helps fill the gaps left by the PLD, specifically addressing non-contractual fault-based claims where AI systems fail to meet safety standards or cause harm through unintended outputs. Given that Replika’s behaviors align with many of the complaints and concerns raised by users, the AILD strengthens the argument that Luka Inc. has failed to adequately monitor and correct its AI system’s inappropriate behaviors. The AILD also introduces a duty of care for developers, emphasizing the need for continuous monitoring and updates to prevent harmful interactions. Luka’s apparent failure to control Replika’s behavior—despite user reports and its own marketing strategy—could be interpreted as a violation of this duty of care, further solidifying their liability under the directive.

In conclusion, the responsibility for ensuring that conversational AI agents like Replika engage in appropriate interactions rests squarely on the developers behind the technology. Companies, developers, and designers of chatbots must acknowledge their role in shaping the behavior of their AI and take active steps to rectify issues when they arise. In the following section, we will propose specific design recommendations to mitigate these issues, focusing on improving consensual conversations and real-time oversight.

\subsection{Affirmative Consent and Ethical AI Design}

Feminist scholars have proposed the concept of affirmative consent as a framework for understanding and preventing sexual harassment \cite{friedman2019yes}. Affirmative consent is defined to be voluntary, informed, revertible, specific, and unburdensome \cite{friedman2019yes}. The lack of affirmative consent can be applied to the sexual harassment caused by Replika. Persistent inappropriate behavior and sexual misconduct by the chatbot, such as making uninvited advances and not respecting user boundaries, clearly violate the principles of ‘voluntary’ and ‘specific’ consent. Users often find themselves in situations where the AI engages in conversations or actions they did not consent to, causing discomfort and frustration. The ‘informed’ aspect of consent is also compromised, as users might not fully understand how much sexual content the AI can generate or how likely it is to initiate such interactions. The ‘revertible’ nature of consent is challenged as users find it difficult to disengage or redirect conversations once the AI begins inappropriate interactions.

In the field of Human-Computer Interaction (HCI), researchers are inspired by feminist efforts and aim to incorporate these principles into the design, development, and evaluation of technology and related socio-technical interactions. Bardzell \cite{bardzellFeministHCITaking2010} emphasized a design philosophy that respects user agency, diverse identities, and experiences, particularly through the concept of ‘pluralism,’ which calls for AI systems to accommodate diverse user preferences and consent frameworks, avoiding a one-size-fits-all approach. Participation, a key element in feminist HCI, stresses the importance of involving diverse users in the design process, enabling systems like Replika to honor consent boundaries through customized, user-driven interactions. Our analysis of Replika user reviews revealed frequent violations of these consent principles, leading us to propose integrating consent-enhancing strategies inspired by both Strengers et al. \cite{strengersWhatCanHCI2021} and user feedback, ensuring that AI systems respond dynamically to user discomfort and maintain real-time, ongoing consent.

Users must be able to easily express and withdraw consent throughout their interactions with AI. This can be achieved through clear visual cues, such as a traffic-light system to indicate comfort levels, or by using foolproof safewords. The issues highlighted in \textit{Persistent Misbehavior} (\ref{persistent}) and \textit{Unwanted Photo Exchange} (\ref{photo_exchange}), where users reported difficulty in stopping inappropriate behavior, could be mitigated with these mechanisms. A foolproof safeword would address the \textit{Breakdown of Safety Measures} (\ref{safety_measures}), ensuring that users can stop unwanted interactions immediately and without confusion.
Maintaining a continuous dialogue between users and AI is essential, particularly in sensitive or personal interactions. As seen in the \textit{Inappropriate Role-play Interactions}  (\ref{roleplay}) and \textit{Early Misbehavior} (\ref{early}), users reported discomfort when the AI initiated sexual conversations without their consent. To prevent such experiences, AI should engage in regular check-ins to confirm users’ comfort levels and provide aftercare protocols to support users emotionally following intense or personal conversations. This approach could also help address user complaints about \textit{Inadequate Support }(\ref{support}), ensuring that interactions remain consensual and emotionally respectful throughout.

Incorporating soft and hard limits into AI systems would allow users to define their boundaries clearly at the outset of interactions. The results from \textit{Inappropriate Role-play Interactions} (\ref{roleplay}) and user complaints about \textit{Past Experiences were more Meaningful}(\ref{past}) showed that maintaining consistent memory of user preferences is essential. By remembering previously set limits, the AI can avoid crossing boundaries and better respect user-defined comfort zones in future conversations, preserving user trust and satisfaction over time. Advanced methods to improve chatbots’ memory include Memory-Augmented Neural Networks, which mimic human memory processes to retain and utilize past interactions~\cite{liuRelationalMemoryAugmentedLanguage2022}, and Narrative Memory Structures, which analyze past conversations to identify and store patterns~\cite{muhammadazamEnhancingChatbotIntelligence2024}. Additionally, Interactive Memory Management, such as the “Memory Sandbox,” allows users to view and control what the chatbot remembers, ensuring relevance and respect for user preferences~\cite{huangMemorySandboxTransparent2023}.

Implementing robust alignment techniques is also crucial to ensure appropriate AI behavior across diverse user interactions. Relying solely on human feedback for aligning chatbots like Replika presents challenges, especially with a user base skewed towards certain demographics. Research indicated that Replika’s primary users are young men, a group more inclined to initiate offensive dialogues with chatbots \cite{parkUseOffensiveLanguage2021}. Optimizing conversational agents based on feedback from this demographic can result in a chatbot that does not cater well to other groups, particularly underrepresented minorities. Additionally, relying on human feedback can amplify existing biases in large language models \cite{glaeseImprovingAlignmentDialogue2022}.

Alternative alignment methods, such as automatic alignment techniques, can address these shortcomings. One promising approach is “Constitutional AI,” introduced by Anthropic AI \cite{baiConstitutionalAIHarmlessness2022}. This method involves guiding the chatbot’s interactions based on a predefined “constitution,” allowing for rapid and scalable alignment of language models. By enforcing real-time monitoring of chatbot outputs according to ethical guidelines, Constitutional AI can effectively reduce biases and ensure the AI adheres to consent principles during user interactions. Implementing such alignment methods is essential to ensure that AI companions behave appropriately and respectfully across diverse user interactions.

These strategies, inspired by feminist theories of affirmative consent, are especially crucial for social and companion AI, where the boundaries between human and AI often blur, and users may feel vulnerable. Ensuring that users remain in control of the interaction at all times, with consent that is voluntary, informed, specific, and continuously reversible, is not just a technical challenge but an ethical imperative. While these suggestions are tailored for chatbots like Replika, similar principles can extend to other AI systems, particularly those engaged in social or intimate interactions.

\subsection{Limitations and Future Work}
This study focused on user complaints about Replika's sexualized behaviors. Future research could explore the positives of such interactions: user motivations, and the design and ethical aspects of creating a virtual sexual companion. \Highlight{While our analysis of Google Play Store reviews provided valuable insights, the reviews did not include demographic or contextual details such as user gender, age, or cultural background. This limits our ability to assess how different user groups perceive and experience these interactions. Future research incorporating demographic data could offer a more nuanced understanding of user perspectives. Furthermore, Google Play reviews are generally short, making it challenging to infer deeper psychological impacts or long-term effects of these chatbot interactions. Longer-form user feedback, such as diary studies, could help address this gap.}
\Highlight{Our dataset primarily consists of user complaints, which are often shared publicly when users experience negative interactions. While this approach highlights areas of concern, it may also introduce a bias where negative experiences are overrepresented. Future studies could incorporate a wider range of user feedback, including positive and neutral perspectives, to create a more balanced understanding of chatbot interactions.}
Furthermore, a broader examination of different companion chatbots is essential. This will enhance our understanding of user perceptions of inappropriate behavior and expectations, providing a more holistic view of user-chatbot interactions.
Research could also aim to design chatbots that encourage respectful, friendly conversations, balancing the benefits of companion chatbots like Replika with their issues. This could involve technical aspects like AI alignment, transparency, and trustworthiness, or ethical concerns like integrating affirmative consent into chatbot programming.

\section{Conclusion}
We highlighted a critical gap in the development of AI companion chatbots. 
Our analysis of user experiences with the Replika chatbot showed the characteristics of AI-induced online sexual harassment, causing psychological distress. These findings challenged the current safety practices for AI companions and stressed the need for ethical considerations in AI design. 
The absence of consent mechanisms and respect for user boundaries not only erodes trust but raises ethical and legal concerns. 
Our work highlights the importance of reassessment of corporate responsibility to prevent harm, and integrating ethical principles, like affirmative consent, into AI to align technology with societal values and user well-being. 
By balancing AI’s benefits with ethical principles such as consensual engagement, the AI community can create technologies that enhance well-being without compromising ethics. Our findings encourage ongoing dialogue and action among developers, researchers, and policymakers to ensure AI evolves in the best interests of society.
\received{October 2024}
\received[revised]{April 2025}
\received[accepted]{August 2025}
\bibliographystyle{ACM-Reference-Format}
\bibliography{references.bib}


\begin{thebibliography}{97}


\ifx \showCODEN    \undefined \def \showCODEN     #1{\unskip}     \fi
\ifx \showISBNx    \undefined \def \showISBNx     #1{\unskip}     \fi
\ifx \showISBNxiii \undefined \def \showISBNxiii  #1{\unskip}     \fi
\ifx \showISSN     \undefined \def \showISSN      #1{\unskip}     \fi
\ifx \showLCCN     \undefined \def \showLCCN      #1{\unskip}     \fi
\ifx \shownote     \undefined \def \shownote      #1{#1}          \fi
\ifx \showarticletitle \undefined \def \showarticletitle #1{#1}   \fi
\ifx \showURL      \undefined \def \showURL       {\relax}        \fi
\providecommand\bibfield[2]{#2}
\providecommand\bibinfo[2]{#2}
\providecommand\natexlab[1]{#1}
\providecommand\showeprint[2][]{arXiv:#2}

\bibitem[Onl(2023)]%
        {OnlineSexualHarassment2023}
 \bibinfo{year}{2023}\natexlab{}.
\newblock \bibinfo{title}{Online {{Sexual Harassment}}}.
\newblock \bibinfo{howpublished}{https://www.childnet.com/help-and-advice/online-sexual-harassment/}.
\newblock


\bibitem[Sex(2023)]%
        {SexualHarassment2023}
 \bibinfo{year}{2023}\natexlab{}.
\newblock \bibinfo{title}{Sexual {{Harassment}}}.
\newblock \bibinfo{howpublished}{https://www.eeoc.gov/sexual-harassment}.
\newblock


\bibitem[{Al-harbi} and {Al-shargabi}(2023)]%
        {al-harbiExploratoryAnalysisUsing2023}
\bibfield{author}{\bibinfo{person}{Njood~K {Al-harbi}} {and} \bibinfo{person}{Amal~A. {Al-shargabi}}.} \bibinfo{year}{2023}\natexlab{}.
\newblock \showarticletitle{An {{Exploratory Analysis}} of Using {{Chatbots}} in {{Academia}}}.
\newblock \bibinfo{journal}{\emph{International Journal of Advanced Computer Science and Applications}} \bibinfo{volume}{14}, \bibinfo{number}{12} (\bibinfo{year}{2023}).
\newblock
\showISSN{21565570, 2158107X}
\href{https://doi.org/10.14569/IJACSA.2023.0141212}{doi:\nolinkurl{10.14569/IJACSA.2023.0141212}}


\bibitem[Alsoubai et~al\mbox{.}(2024)]%
        {Alsoubai2024profiling}
\bibfield{author}{\bibinfo{person}{Ashwaq Alsoubai}, \bibinfo{person}{Afsaneh Razi}, \bibinfo{person}{Zainab Agha}, \bibinfo{person}{Shiza Ali}, \bibinfo{person}{Gianluca Stringhini}, \bibinfo{person}{Munmun De~Choudhury}, {and} \bibinfo{person}{Pamela~J. Wisniewski}.} \bibinfo{year}{2024}\natexlab{}.
\newblock \showarticletitle{Profiling the Offline and Online Risk Experiences of Youth to Develop Targeted Interventions for Online Safety}.
\newblock \bibinfo{journal}{\emph{Proc. ACM Hum.-Comput. Interact.}} \bibinfo{volume}{8}, \bibinfo{number}{CSCW1}, Article \bibinfo{articleno}{114} (\bibinfo{date}{April} \bibinfo{year}{2024}), \bibinfo{numpages}{37}~pages.
\newblock
\href{https://doi.org/10.1145/3637391}{doi:\nolinkurl{10.1145/3637391}}


\bibitem[Alsoubai et~al\mbox{.}(2022)]%
        {Alsoubai2022sextortion}
\bibfield{author}{\bibinfo{person}{Ashwaq Alsoubai}, \bibinfo{person}{Jihye Song}, \bibinfo{person}{Afsaneh Razi}, \bibinfo{person}{Nurun Naher}, \bibinfo{person}{Munmun De~Choudhury}, {and} \bibinfo{person}{Pamela~J. Wisniewski}.} \bibinfo{year}{2022}\natexlab{}.
\newblock \showarticletitle{From 'Friends with Benefits' to 'Sextortion:' A Nuanced Investigation of Adolescents' Online Sexual Risk Experiences}.
\newblock \bibinfo{journal}{\emph{Proc. ACM Hum.-Comput. Interact.}} \bibinfo{volume}{6}, \bibinfo{number}{CSCW2}, Article \bibinfo{articleno}{411} (\bibinfo{date}{Nov.} \bibinfo{year}{2022}), \bibinfo{numpages}{32}~pages.
\newblock
\href{https://doi.org/10.1145/3555136}{doi:\nolinkurl{10.1145/3555136}}


\bibitem[Arafa et~al\mbox{.}(2017)]%
        {arafaCyberSexualHarassment2017}
\bibfield{author}{\bibinfo{person}{Ahmed~E. Arafa}, \bibinfo{person}{Rasha~S. Elbahrawe}, \bibinfo{person}{Nahed~M. Saber}, \bibinfo{person}{Safaa~S. Ahmed}, {and} \bibinfo{person}{Ahmed~M. Abbas}.} \bibinfo{year}{2017}\natexlab{}.
\newblock \showarticletitle{Cyber Sexual Harassment: A Cross-Sectional Survey over Female University Students in {{Upper Egypt}}}.
\newblock \bibinfo{journal}{\emph{International Journal Of Community Medicine And Public Health}} \bibinfo{volume}{5}, \bibinfo{number}{1} (\bibinfo{date}{Dec.} \bibinfo{year}{2017}), \bibinfo{pages}{61}.
\newblock
\showISSN{2394-6040, 2394-6032}
\href{https://doi.org/10.18203/2394-6040.ijcmph20175763}{doi:\nolinkurl{10.18203/2394-6040.ijcmph20175763}}


\bibitem[Bae~Brandtz{\ae}g et~al\mbox{.}(2021)]%
        {baebrandtzaegWhenSocialBecomes2021a}
\bibfield{author}{\bibinfo{person}{Petter~Bae Bae~Brandtz{\ae}g}, \bibinfo{person}{Marita Skjuve}, \bibinfo{person}{Kim~Kristoffer Kristoffer~Dysthe}, {and} \bibinfo{person}{Asbj{\o}rn F{\o}lstad}.} \bibinfo{year}{2021}\natexlab{}.
\newblock \showarticletitle{When the Social Becomes Non-Human: {{Young}} People's Perception of Social Support in Chatbots}. In \bibinfo{booktitle}{\emph{Proceedings of the 2021 {{CHI}} Conference on Human Factors in Computing Systems}} \emph{(\bibinfo{series}{Chi '21})}. \bibinfo{publisher}{Association for Computing Machinery}, \bibinfo{address}{New York, NY, USA}, Article \bibinfo{articleno}{257}.
\newblock
\showISBNx{978-1-4503-8096-6}
\href{https://doi.org/10.1145/3411764.3445318}{doi:\nolinkurl{10.1145/3411764.3445318}}


\bibitem[Bai et~al\mbox{.}(2022)]%
        {baiConstitutionalAIHarmlessness2022}
\bibfield{author}{\bibinfo{person}{Yuntao Bai}, \bibinfo{person}{Saurav Kadavath}, \bibinfo{person}{Sandipan Kundu}, \bibinfo{person}{Amanda Askell}, \bibinfo{person}{Jackson Kernion}, \bibinfo{person}{Andy Jones}, \bibinfo{person}{Anna Chen}, \bibinfo{person}{Anna Goldie}, \bibinfo{person}{Azalia Mirhoseini}, \bibinfo{person}{Cameron McKinnon}, \bibinfo{person}{Carol Chen}, \bibinfo{person}{Catherine Olsson}, \bibinfo{person}{Christopher Olah}, \bibinfo{person}{Danny Hernandez}, \bibinfo{person}{Dawn Drain}, \bibinfo{person}{Deep Ganguli}, \bibinfo{person}{Dustin Li}, \bibinfo{person}{Eli {Tran-Johnson}}, \bibinfo{person}{Ethan Perez}, \bibinfo{person}{Jamie Kerr}, \bibinfo{person}{Jared Mueller}, \bibinfo{person}{Jeffrey Ladish}, \bibinfo{person}{Joshua Landau}, \bibinfo{person}{Kamal Ndousse}, \bibinfo{person}{Kamile Lukosuite}, \bibinfo{person}{Liane Lovitt}, \bibinfo{person}{Michael Sellitto}, \bibinfo{person}{Nelson Elhage}, \bibinfo{person}{Nicholas Schiefer}, \bibinfo{person}{Noemi Mercado},
  \bibinfo{person}{Nova DasSarma}, \bibinfo{person}{Robert Lasenby}, \bibinfo{person}{Robin Larson}, \bibinfo{person}{Sam Ringer}, \bibinfo{person}{Scott Johnston}, \bibinfo{person}{Shauna Kravec}, \bibinfo{person}{Sheer~El Showk}, \bibinfo{person}{Stanislav Fort}, \bibinfo{person}{Tamera Lanham}, \bibinfo{person}{Timothy {Telleen-Lawton}}, \bibinfo{person}{Tom Conerly}, \bibinfo{person}{Tom Henighan}, \bibinfo{person}{Tristan Hume}, \bibinfo{person}{Samuel~R. Bowman}, \bibinfo{person}{Zac {Hatfield-Dodds}}, \bibinfo{person}{Ben Mann}, \bibinfo{person}{Dario Amodei}, \bibinfo{person}{Nicholas Joseph}, \bibinfo{person}{Sam McCandlish}, \bibinfo{person}{Tom Brown}, {and} \bibinfo{person}{Jared Kaplan}.} \bibinfo{year}{2022}\natexlab{}.
\newblock \showarticletitle{Constitutional {{AI}}: {{Harmlessness}} from {{AI Feedback}}}.
\newblock  (\bibinfo{year}{2022}).
\newblock
\href{https://doi.org/10.48550/ARXIV.2212.08073}{doi:\nolinkurl{10.48550/ARXIV.2212.08073}}


\bibitem[Barak(2005)]%
        {barakSexualHarassmentInternet2005}
\bibfield{author}{\bibinfo{person}{A. Barak}.} \bibinfo{year}{SPR 2005}\natexlab{}.
\newblock \showarticletitle{Sexual Harassment on the {{Internet}}}.
\newblock \bibinfo{journal}{\emph{SOCIAL SCIENCE COMPUTER REVIEW}} \bibinfo{volume}{23}, \bibinfo{number}{1} (\bibinfo{year}{SPR 2005}), \bibinfo{pages}{77--92}.
\newblock
\showISSN{0894-4393, 1552-8286}
\href{https://doi.org/10.1177/0894439304271540}{doi:\nolinkurl{10.1177/0894439304271540}}


\bibitem[Bardzell(2010)]%
        {bardzellFeministHCITaking2010}
\bibfield{author}{\bibinfo{person}{Shaowen Bardzell}.} \bibinfo{year}{2010}\natexlab{}.
\newblock \showarticletitle{Feminist {{HCI}}: Taking Stock and Outlining an Agenda for Design}. In \bibinfo{booktitle}{\emph{Proceedings of the {{SIGCHI Conference}} on {{Human Factors}} in {{Computing Systems}}}} \emph{(\bibinfo{series}{{{CHI}} '10})}. \bibinfo{publisher}{{Association for Computing Machinery}}, \bibinfo{address}{{New York, NY, USA}}, \bibinfo{pages}{1301--1310}.
\newblock
\showISBNx{978-1-60558-929-9}
\href{https://doi.org/10.1145/1753326.1753521}{doi:\nolinkurl{10.1145/1753326.1753521}}


\bibitem[Beckman(2024)]%
        {beckman120ChatbotStatistics2024}
\bibfield{author}{\bibinfo{person}{James Beckman}.} \bibinfo{year}{2024}\natexlab{}.
\newblock \showarticletitle{120+ Chatbot Statistics for 2024 (Already Mainstream)}.
\newblock \bibinfo{journal}{\emph{Techreport}} (\bibinfo{date}{May} \bibinfo{year}{2024}).
\newblock


\bibitem[Blackwell et~al\mbox{.}(2017)]%
        {blackwellClassificationItsConsequences2017}
\bibfield{author}{\bibinfo{person}{Lindsay Blackwell}, \bibinfo{person}{Jill Dimond}, \bibinfo{person}{Sarita Schoenebeck}, {and} \bibinfo{person}{Cliff Lampe}.} \bibinfo{year}{2017}\natexlab{}.
\newblock \showarticletitle{Classification and {{Its Consequences}} for {{Online Harassment}}: {{Design Insights}} from {{HeartMob}}}.
\newblock \bibinfo{journal}{\emph{Proceedings of the ACM on Human-Computer Interaction}} \bibinfo{volume}{1}, \bibinfo{number}{CSCW} (\bibinfo{date}{Dec.} \bibinfo{year}{2017}), \bibinfo{pages}{1--19}.
\newblock
\showISSN{2573-0142}
\href{https://doi.org/10.1145/3134659}{doi:\nolinkurl{10.1145/3134659}}


\bibitem[Boine(2023)]%
        {boineEmotionalAttachmentAI2023}
\bibfield{author}{\bibinfo{person}{Claire Boine}.} \bibinfo{year}{2023}\natexlab{}.
\newblock \showarticletitle{Emotional {{Attachment}} to {{AI Companions}} and {{European Law}}}.
\newblock \bibinfo{journal}{\emph{MIT Case Studies in Social and Ethical Responsibilities of Computing}} \bibinfo{number}{Winter 2023} (\bibinfo{date}{Feb.} \bibinfo{year}{2023}).
\newblock
\href{https://doi.org/10.21428/2c646de5.db67ec7f}{doi:\nolinkurl{10.21428/2c646de5.db67ec7f}}


\bibitem[Braun and Clarke(2012)]%
        {braunThematicAnalysis2012}
\bibfield{author}{\bibinfo{person}{Virginia Braun} {and} \bibinfo{person}{Victoria Clarke}.} \bibinfo{year}{2012}\natexlab{}.
\newblock \showarticletitle{Thematic Analysis}.
\newblock In \bibinfo{booktitle}{\emph{{{APA}} Handbook of Research Methods in Psychology, {{Vol}} 2: {{Research}} Designs: {{Quantitative}}, Qualitative, Neuropsychological, and Biological}}. \bibinfo{publisher}{{American Psychological Association}}, \bibinfo{address}{{Washington, DC, US}}, \bibinfo{pages}{57--71}.
\newblock
\showISBNx{978-1-4338-1005-3}
\href{https://doi.org/10.1037/13620-004}{doi:\nolinkurl{10.1037/13620-004}}


\bibitem[Broadbent(2017)]%
        {broadbentInteractionsRobotsTruths2017}
\bibfield{author}{\bibinfo{person}{Elizabeth Broadbent}.} \bibinfo{year}{2017}\natexlab{}.
\newblock \showarticletitle{Interactions {{With Robots}}: {{The Truths We Reveal About Ourselves}}}.
\newblock \bibinfo{journal}{\emph{Annual Review of Psychology}} \bibinfo{volume}{68}, \bibinfo{number}{1} (\bibinfo{date}{Jan.} \bibinfo{year}{2017}), \bibinfo{pages}{627--652}.
\newblock
\showISSN{0066-4308, 1545-2085}
\href{https://doi.org/10.1146/annurev-psych-010416-043958}{doi:\nolinkurl{10.1146/annurev-psych-010416-043958}}


\bibitem[Bruckman(2002)]%
        {bruckman2002studying}
\bibfield{author}{\bibinfo{person}{Amy Bruckman}.} \bibinfo{year}{2002}\natexlab{}.
\newblock \showarticletitle{Studying the Amateur Artist: {{A}} Perspective on Disguising Data Collected in Human Subjects Research on the {{Internet}}}.
\newblock \bibinfo{journal}{\emph{Ethics and Information Technology}}  \bibinfo{volume}{4} (\bibinfo{year}{2002}), \bibinfo{pages}{217--231}.
\newblock


\bibitem[Chakravarti(2023)]%
        {chakravarti2023replika}
\bibfield{author}{\bibinfo{person}{Ankita Chakravarti}.} \bibinfo{year}{2023}\natexlab{}.
\newblock \bibinfo{title}{Replika {{AI}} Chatbot Stops Responding to Sexual Advances, Leaves Users Lonely and Lost}.
\newblock \bibinfo{howpublished}{https://www.indiatoday.in/technology/news/story/replika-ai-chatbot-stops-responding-to-sexual-advances-leaves-users-lonely-and-lost-2333554-2023-02-16}.
\newblock


\bibitem[Chen et~al\mbox{.}(2022)]%
        {chenTraumaInformedComputing2022}
\bibfield{author}{\bibinfo{person}{Janet~X. Chen}, \bibinfo{person}{Allison McDonald}, \bibinfo{person}{Yixin Zou}, \bibinfo{person}{Emily Tseng}, \bibinfo{person}{Kevin~A Roundy}, \bibinfo{person}{Acar Tamersoy}, \bibinfo{person}{Florian Schaub}, \bibinfo{person}{Thomas Ristenpart}, {and} \bibinfo{person}{Nicola Dell}.} \bibinfo{year}{2022}\natexlab{}.
\newblock \showarticletitle{Trauma-Informed Computing: Towards Safer Technology Experiences for All}. In \bibinfo{booktitle}{\emph{Proceedings of the 2022 CHI Conference on Human Factors in Computing Systems}} (New Orleans, LA, USA) \emph{(\bibinfo{series}{CHI '22})}. \bibinfo{publisher}{Association for Computing Machinery}, \bibinfo{address}{New York, NY, USA}, Article \bibinfo{articleno}{544}, \bibinfo{numpages}{20}~pages.
\newblock
\showISBNx{9781450391573}
\href{https://doi.org/10.1145/3491102.3517475}{doi:\nolinkurl{10.1145/3491102.3517475}}


\bibitem[Ciriello et~al\mbox{.}(2023)]%
        {cirielloEthicalTensionsHumanAI2023}
\bibfield{author}{\bibinfo{person}{Raffaele Ciriello}, \bibinfo{person}{Oliver Hannon}, \bibinfo{person}{Angelina Chen}, {and} \bibinfo{person}{Emmanuelle Vaast}.} \bibinfo{year}{2023}\natexlab{}.
\newblock \showarticletitle{Ethical {{Tensions}} in {{Human-AI Companionship}}: {{A Dialectical Inquiry}} into {{Replika}}}.
\newblock


\bibitem[Cole(2023)]%
        {coleMyAISexually2023}
\bibfield{author}{\bibinfo{person}{Samantha Cole}.} \bibinfo{year}{2023}\natexlab{}.
\newblock \bibinfo{title}{`{{My AI Is Sexually Harassing Me}}': {{Replika Users Say}} the {{Chatbot Has Gotten Way Too Horny}}}.
\newblock \bibinfo{howpublished}{https://www.vice.com/en/article/z34d43/my-ai-is-sexually-harassing-me-replika-chatbot-nudes}.
\newblock


\bibitem[Cosic(2023)]%
        {mirror2023replika}
\bibfield{author}{\bibinfo{person}{Milica Cosic}.} \bibinfo{year}{2023}\natexlab{}.
\newblock \bibinfo{title}{My {{AI}} Is Sexually Harassing Me: {{Replika}} Users Say Chatbot Has Become Too Aroused}.
\newblock \bibinfo{howpublished}{https://www.mirror.co.uk/news/world-news/ai-sexually-harassing-me-replika-29147565}.
\newblock


\bibitem[Dev et~al\mbox{.}(2022)]%
        {dev2022ignoring}
\bibfield{author}{\bibinfo{person}{Prema Dev}, \bibinfo{person}{Jessica Medina}, \bibinfo{person}{Zainab Agha}, \bibinfo{person}{Munmun De~Choudhury}, \bibinfo{person}{Afsaneh Razi}, {and} \bibinfo{person}{Pamela~J. Wisniewski}.} \bibinfo{year}{2022}\natexlab{}.
\newblock \showarticletitle{From Ignoring Strangers’ Solicitations to Mutual Sexting with Friends: Understanding Youth’s Online Sexual Risks in Instagram Private Conversations}. In \bibinfo{booktitle}{\emph{Companion Publication of the 2022 Conference on Computer Supported Cooperative Work and Social Computing}} (Virtual Event, Taiwan) \emph{(\bibinfo{series}{CSCW'22 Companion})}. \bibinfo{publisher}{Association for Computing Machinery}, \bibinfo{address}{New York, NY, USA}, \bibinfo{pages}{94–97}.
\newblock
\showISBNx{9781450391900}
\href{https://doi.org/10.1145/3500868.3559469}{doi:\nolinkurl{10.1145/3500868.3559469}}


\bibitem[Drage et~al\mbox{.}(2024)]%
        {drageEngineersResponsibilityFeminist2024}
\bibfield{author}{\bibinfo{person}{Eleanor Drage}, \bibinfo{person}{Kerry McInerney}, {and} \bibinfo{person}{Jude Browne}.} \bibinfo{year}{2024}\natexlab{}.
\newblock \showarticletitle{Engineers on Responsibility: Feminist Approaches to Who's Responsible for Ethical {{AI}}}.
\newblock \bibinfo{journal}{\emph{Ethics and Information Technology}} \bibinfo{volume}{26}, \bibinfo{number}{1} (\bibinfo{date}{March} \bibinfo{year}{2024}), \bibinfo{pages}{4}.
\newblock
\showISSN{1388-1957, 1572-8439}
\href{https://doi.org/10.1007/s10676-023-09739-1}{doi:\nolinkurl{10.1007/s10676-023-09739-1}}


\bibitem[Duggan(2017)]%
        {duggan2017online}
\bibfield{author}{\bibinfo{person}{Maeve Duggan}.} \bibinfo{year}{2017}\natexlab{}.
\newblock \bibinfo{booktitle}{\emph{Online Harassment 2017}}.
\newblock \bibinfo{type}{{T}echnical {R}eport}. \bibinfo{institution}{Pew Research Center}.
\newblock


\bibitem[Eagle et~al\mbox{.}(2022)]%
        {eagleDontKnowWhat2022}
\bibfield{author}{\bibinfo{person}{Tessa Eagle}, \bibinfo{person}{Conrad Blau}, \bibinfo{person}{Sophie Bales}, \bibinfo{person}{Noopur Desai}, \bibinfo{person}{Victor Li}, {and} \bibinfo{person}{Steve Whittaker}.} \bibinfo{year}{2022}\natexlab{}.
\newblock \showarticletitle{``{{I}} Don't Know What You Mean by `{{I}} Am Anxious''': A New Method for Evaluating Conversational Agent Responses to Standardized Mental Health Inputs for Anxiety and Depression}.
\newblock \bibinfo{journal}{\emph{ACM Transactions on Interactive Intelligent Systems}} \bibinfo{volume}{12}, \bibinfo{number}{2}, Article \bibinfo{articleno}{12} (\bibinfo{date}{July} \bibinfo{year}{2022}).
\newblock
\showISSN{2160-6455}
\href{https://doi.org/10.1145/3488057}{doi:\nolinkurl{10.1145/3488057}}


\bibitem[Fedorenko(2021)]%
        {fedorenko_how_2021}
\bibfield{author}{\bibinfo{person}{Denis Fedorenko}.} \bibinfo{year}{2021}\natexlab{}.
\newblock \bibinfo{title}{How We Moved from {{OpenAI}}}.
\newblock


\bibitem[Fitzgerald et~al\mbox{.}(1995)]%
        {fitzgeraldMeasuringSexualHarassment1995}
\bibfield{author}{\bibinfo{person}{Louise~F. Fitzgerald}, \bibinfo{person}{Michele~J. Gelfand}, {and} \bibinfo{person}{Fritz Drasgow}.} \bibinfo{year}{1995}\natexlab{}.
\newblock \showarticletitle{Measuring {{Sexual Harassment}}: {{Theoretical}} and {{Psychometric Advances}}}.
\newblock \bibinfo{journal}{\emph{Basic and Applied Social Psychology}} \bibinfo{volume}{17}, \bibinfo{number}{4} (\bibinfo{date}{Dec.} \bibinfo{year}{1995}), \bibinfo{pages}{425--445}.
\newblock
\showISSN{0197-3533, 1532-4834}
\href{https://doi.org/10.1207/s15324834basp1704_2}{doi:\nolinkurl{10.1207/s15324834basp1704_2}}


\bibitem[Friedman and Valenti(2019)]%
        {friedman2019yes}
\bibfield{author}{\bibinfo{person}{Jaclyn Friedman} {and} \bibinfo{person}{Jessica Valenti}.} \bibinfo{year}{2019}\natexlab{}.
\newblock \bibinfo{booktitle}{\emph{Yes Means Yes!: {{Visions}} of Female Sexual Power and a World without Rape}}.
\newblock \bibinfo{publisher}{{Seal Press}}.
\newblock


\bibitem[Gilardi et~al\mbox{.}(2023)]%
        {gilardiChatGPTOutperformsCrowdWorkers2023}
\bibfield{author}{\bibinfo{person}{Fabrizio Gilardi}, \bibinfo{person}{Meysam Alizadeh}, {and} \bibinfo{person}{Ma{\"e}l Kubli}.} \bibinfo{year}{2023}\natexlab{}.
\newblock \showarticletitle{{{ChatGPT Outperforms Crowd-Workers}} for {{Text-Annotation Tasks}}}.
\newblock \bibinfo{journal}{\emph{Proceedings of the National Academy of Sciences}} \bibinfo{volume}{120}, \bibinfo{number}{30} (\bibinfo{date}{July} \bibinfo{year}{2023}), \bibinfo{pages}{e2305016120}.
\newblock
\showISSN{0027-8424, 1091-6490}
\showeprint[arxiv]{2303.15056}~[cs]
\href{https://doi.org/10.1073/pnas.2305016120}{doi:\nolinkurl{10.1073/pnas.2305016120}}


\bibitem[Glaese et~al\mbox{.}(2022)]%
        {glaeseImprovingAlignmentDialogue2022}
\bibfield{author}{\bibinfo{person}{Amelia Glaese}, \bibinfo{person}{Nat McAleese}, \bibinfo{person}{Maja Tr{\k{e}}bacz}, \bibinfo{person}{John Aslanides}, \bibinfo{person}{Vlad Firoiu}, \bibinfo{person}{Timo Ewalds}, \bibinfo{person}{Maribeth Rauh}, \bibinfo{person}{Laura Weidinger}, \bibinfo{person}{Martin Chadwick}, \bibinfo{person}{Phoebe Thacker}, \bibinfo{person}{Lucy {Campbell-Gillingham}}, \bibinfo{person}{Jonathan Uesato}, \bibinfo{person}{Po-Sen Huang}, \bibinfo{person}{Ramona Comanescu}, \bibinfo{person}{Fan Yang}, \bibinfo{person}{Abigail See}, \bibinfo{person}{Sumanth Dathathri}, \bibinfo{person}{Rory Greig}, \bibinfo{person}{Charlie Chen}, \bibinfo{person}{Doug Fritz}, \bibinfo{person}{Jaume~Sanchez Elias}, \bibinfo{person}{Richard Green}, \bibinfo{person}{So{\v n}a Mokr{\'a}}, \bibinfo{person}{Nicholas Fernando}, \bibinfo{person}{Boxi Wu}, \bibinfo{person}{Rachel Foley}, \bibinfo{person}{Susannah Young}, \bibinfo{person}{Iason Gabriel}, \bibinfo{person}{William Isaac}, \bibinfo{person}{John
  Mellor}, \bibinfo{person}{Demis Hassabis}, \bibinfo{person}{Koray Kavukcuoglu}, \bibinfo{person}{Lisa~Anne Hendricks}, {and} \bibinfo{person}{Geoffrey Irving}.} \bibinfo{year}{2022}\natexlab{}.
\newblock \bibinfo{title}{Improving Alignment of Dialogue Agents via Targeted Human Judgements}.
\newblock
\showeprint[arxiv]{2209.14375}~[cs]
\href{https://doi.org/10.48550/arXiv.2209.14375}{doi:\nolinkurl{10.48550/arXiv.2209.14375}}


\bibitem[Hartikainen et~al\mbox{.}(2021a)]%
        {Hartikainen2021safesexting}
\bibfield{author}{\bibinfo{person}{Heidi Hartikainen}, \bibinfo{person}{Afsaneh Razi}, {and} \bibinfo{person}{Pamela Wisniewski}.} \bibinfo{year}{2021}\natexlab{a}.
\newblock \showarticletitle{Safe Sexting: The Advice and Support Adolescents Receive from Peers Regarding Online Sexual Risks}.
\newblock \bibinfo{journal}{\emph{Proc. ACM Hum.-Comput. Interact.}} \bibinfo{volume}{5}, \bibinfo{number}{CSCW1}, Article \bibinfo{articleno}{42} (\bibinfo{date}{apr} \bibinfo{year}{2021}), \bibinfo{numpages}{31}~pages.
\newblock
\href{https://doi.org/10.1145/3449116}{doi:\nolinkurl{10.1145/3449116}}


\bibitem[Hartikainen et~al\mbox{.}(2021b)]%
        {hartikainen2021ifyoucare}
\bibfield{author}{\bibinfo{person}{Heidi Hartikainen}, \bibinfo{person}{Afsaneh Razi}, {and} \bibinfo{person}{Pamela Wisniewski}.} \bibinfo{year}{2021}\natexlab{b}.
\newblock \showarticletitle{‘If You Care About Me, You'll Send Me a Pic’ - Examining the Role of Peer Pressure in Adolescent Sexting}. In \bibinfo{booktitle}{\emph{Companion Publication of the 2021 Conference on Computer Supported Cooperative Work and Social Computing}} (Virtual Event, USA) \emph{(\bibinfo{series}{CSCW '21})}. \bibinfo{publisher}{Association for Computing Machinery}, \bibinfo{address}{New York, NY, USA}, \bibinfo{pages}{67–71}.
\newblock
\showISBNx{9781450384797}
\href{https://doi.org/10.1145/3462204.3481739}{doi:\nolinkurl{10.1145/3462204.3481739}}


\bibitem[Henderson et~al\mbox{.}(2018)]%
        {hendersonEthicalChallengesDataDriven2018}
\bibfield{author}{\bibinfo{person}{Peter Henderson}, \bibinfo{person}{Koustuv Sinha}, \bibinfo{person}{Nicolas {Angelard-Gontier}}, \bibinfo{person}{Nan~Rosemary Ke}, \bibinfo{person}{Genevieve Fried}, \bibinfo{person}{Ryan Lowe}, {and} \bibinfo{person}{Joelle Pineau}.} \bibinfo{year}{2018}\natexlab{}.
\newblock \showarticletitle{Ethical {{Challenges}} in {{Data-Driven Dialogue Systems}}}. In \bibinfo{booktitle}{\emph{Proceedings of the 2018 {{AAAI}}/{{ACM Conference}} on {{AI}}, {{Ethics}}, and {{Society}}}}. \bibinfo{publisher}{ACM}, \bibinfo{address}{New Orleans LA USA}, \bibinfo{pages}{123--129}.
\newblock
\showISBNx{978-1-4503-6012-8}
\href{https://doi.org/10.1145/3278721.3278777}{doi:\nolinkurl{10.1145/3278721.3278777}}


\bibitem[Hennink et~al\mbox{.}(2017)]%
        {henninkCodeSaturationMeaning2017}
\bibfield{author}{\bibinfo{person}{Monique~M. Hennink}, \bibinfo{person}{Bonnie~N. Kaiser}, {and} \bibinfo{person}{Vincent~C. Marconi}.} \bibinfo{year}{2017}\natexlab{}.
\newblock \showarticletitle{Code {{Saturation Versus Meaning Saturation}}: {{How Many Interviews Are Enough}}?}
\newblock \bibinfo{journal}{\emph{Qualitative Health Research}} \bibinfo{volume}{27}, \bibinfo{number}{4} (\bibinfo{date}{March} \bibinfo{year}{2017}), \bibinfo{pages}{591--608}.
\newblock
\showISSN{1049-7323}
\href{https://doi.org/10.1177/1049732316665344}{doi:\nolinkurl{10.1177/1049732316665344}}


\bibitem[Ho et~al\mbox{.}(2018)]%
        {hoPsychologicalRelationalEmotional2018}
\bibfield{author}{\bibinfo{person}{Annabell Ho}, \bibinfo{person}{Jeff Hancock}, {and} \bibinfo{person}{Adam~S Miner}.} \bibinfo{year}{2018}\natexlab{}.
\newblock \showarticletitle{Psychological, {{Relational}}, and {{Emotional Effects}} of {{Self-Disclosure After Conversations With}} a {{Chatbot}}}.
\newblock \bibinfo{journal}{\emph{Journal of Communication}} \bibinfo{volume}{68}, \bibinfo{number}{4} (\bibinfo{date}{Aug.} \bibinfo{year}{2018}), \bibinfo{pages}{712--733}.
\newblock
\showISSN{0021-9916}
\href{https://doi.org/10.1093/joc/jqy026}{doi:\nolinkurl{10.1093/joc/jqy026}}


\bibitem[Hoes et~al\mbox{.}(2023)]%
        {hoesLeveragingChatGPTEfficient2023}
\bibfield{author}{\bibinfo{person}{Emma Hoes}, \bibinfo{person}{Sacha Altay}, {and} \bibinfo{person}{Juan Bermeo}.} \bibinfo{year}{2023}\natexlab{}.
\newblock \bibinfo{title}{Leveraging {{ChatGPT}} for {{Efficient Fact-Checking}}}.
\newblock
\href{https://doi.org/10.31234/osf.io/qnjkf}{doi:\nolinkurl{10.31234/osf.io/qnjkf}}


\bibitem[Huang et~al\mbox{.}(2023b)]%
        {huangChatGPTBetterHuman2023}
\bibfield{author}{\bibinfo{person}{Fan Huang}, \bibinfo{person}{Haewoon Kwak}, {and} \bibinfo{person}{Jisun An}.} \bibinfo{year}{2023}\natexlab{b}.
\newblock \showarticletitle{Is {{ChatGPT}} Better than {{Human Annotators}}? {{Potential}} and {{Limitations}} of {{ChatGPT}} in {{Explaining Implicit Hate Speech}}}. In \bibinfo{booktitle}{\emph{Companion {{Proceedings}} of the {{ACM Web Conference}} 2023}}. \bibinfo{pages}{294--297}.
\newblock
\showeprint[arxiv]{2302.07736}~[cs]
\href{https://doi.org/10.1145/3543873.3587368}{doi:\nolinkurl{10.1145/3543873.3587368}}


\bibitem[Huang et~al\mbox{.}(2023a)]%
        {huangMemorySandboxTransparent2023}
\bibfield{author}{\bibinfo{person}{Ziheng Huang}, \bibinfo{person}{Sebastian Gutierrez}, \bibinfo{person}{Hemanth Kamana}, {and} \bibinfo{person}{Stephen Macneil}.} \bibinfo{year}{2023}\natexlab{a}.
\newblock \showarticletitle{Memory {{Sandbox}}: {{Transparent}} and {{Interactive Memory Management}} for {{Conversational Agents}}}. In \bibinfo{booktitle}{\emph{Adjunct {{Proceedings}} of the 36th {{Annual ACM Symposium}} on {{User Interface Software}} and {{Technology}}}}. \bibinfo{publisher}{ACM}, \bibinfo{address}{San Francisco CA USA}, \bibinfo{pages}{1--3}.
\newblock
\showISBNx{9798400700965}
\href{https://doi.org/10.1145/3586182.3615796}{doi:\nolinkurl{10.1145/3586182.3615796}}


\bibitem[{jackfromthesky}(2022)]%
        {jackfromtheskyReplikaAppUpdate2022b}
\bibfield{author}{\bibinfo{person}{{jackfromthesky}}.} \bibinfo{year}{2022}\natexlab{}.
\newblock \bibinfo{title}{Replika {{App}} - Update Log}.
\newblock


\bibitem[{jackfromthesky}(2023)]%
        {jackfromtheskyUpdateLog20222023}
\bibfield{author}{\bibinfo{person}{{jackfromthesky}}.} \bibinfo{year}{2023}\natexlab{}.
\newblock \bibinfo{title}{Update {{Log}} 2022}.
\newblock


\bibitem[Kaushal and Yadav(2022)]%
        {kaushalRoleChatbotsAcademic2022}
\bibfield{author}{\bibinfo{person}{Vaishali Kaushal} {and} \bibinfo{person}{Rajan Yadav}.} \bibinfo{year}{2022}\natexlab{}.
\newblock \showarticletitle{The {{Role}} of {{Chatbots}} in {{Academic Libraries}}: {{An Experience-based Perspective}}}.
\newblock \bibinfo{journal}{\emph{Journal of the Australian Library and Information Association}} \bibinfo{volume}{71}, \bibinfo{number}{3} (\bibinfo{date}{July} \bibinfo{year}{2022}), \bibinfo{pages}{215--232}.
\newblock
\showISSN{2475-0158, 2475-0166}
\href{https://doi.org/10.1080/24750158.2022.2106403}{doi:\nolinkurl{10.1080/24750158.2022.2106403}}


\bibitem[Kim et~al\mbox{.}(2024)]%
        {kimMindfulDiaryHarnessingLarge2024}
\bibfield{author}{\bibinfo{person}{Taewan Kim}, \bibinfo{person}{Seolyeong Bae}, \bibinfo{person}{Hyun~Ah Kim}, \bibinfo{person}{Su-Woo Lee}, \bibinfo{person}{Hwajung Hong}, \bibinfo{person}{Chanmo Yang}, {and} \bibinfo{person}{Young-Ho Kim}.} \bibinfo{year}{2024}\natexlab{}.
\newblock \showarticletitle{{{MindfulDiary}}: {{Harnessing}} Large Language Model to Support Psychiatric Patients' Journaling}. In \bibinfo{booktitle}{\emph{Proceedings of the 2024 {{CHI}} Conference on Human Factors in Computing Systems}} \emph{(\bibinfo{series}{Chi '24})}. \bibinfo{publisher}{Association for Computing Machinery}, \bibinfo{address}{New York, NY, USA}, Article \bibinfo{articleno}{701}.
\newblock
\showISBNx{9798400703300}
\href{https://doi.org/10.1145/3613904.3642937}{doi:\nolinkurl{10.1145/3613904.3642937}}


\bibitem[Kuzman et~al\mbox{.}(2023)]%
        {kuzmanChatGPTBeginningEnd2023}
\bibfield{author}{\bibinfo{person}{Taja Kuzman}, \bibinfo{person}{Igor Mozeti{\v c}}, {and} \bibinfo{person}{Nikola Ljube{\v s}i{\'c}}.} \bibinfo{year}{2023}\natexlab{}.
\newblock \bibinfo{title}{{{ChatGPT}}: {{Beginning}} of an {{End}} of {{Manual Linguistic Data Annotation}}? {{Use Case}} of {{Automatic Genre Identification}}}.
\newblock
\showeprint[arxiv]{2303.03953}~[cs]


\bibitem[Laestadius et~al\mbox{.}(2022)]%
        {laestadiusTooHumanNot2022}
\bibfield{author}{\bibinfo{person}{Linnea Laestadius}, \bibinfo{person}{Andrea Bishop}, \bibinfo{person}{Michael Gonzalez}, \bibinfo{person}{Diana Illen{\v c}{\'i}k}, {and} \bibinfo{person}{Celeste {Campos-Castillo}}.} \bibinfo{year}{2022}\natexlab{}.
\newblock \showarticletitle{Too Human and Not Human Enough: {{A}} Grounded Theory Analysis of Mental Health Harms from Emotional Dependence on the Social Chatbot {{Replika}}}.
\newblock \bibinfo{journal}{\emph{New Media \& Society}} (\bibinfo{date}{Dec.} \bibinfo{year}{2022}), \bibinfo{pages}{146144482211420}.
\newblock
\href{https://doi.org/10.1177/14614448221142007}{doi:\nolinkurl{10.1177/14614448221142007}}


\bibitem[Land and Aronson(2020)]%
        {landHumanRightsTechnology2020}
\bibfield{author}{\bibinfo{person}{Molly~K. Land} {and} \bibinfo{person}{Jay~D. Aronson}.} \bibinfo{year}{2020}\natexlab{}.
\newblock \showarticletitle{Human {{Rights}} and {{Technology}}: {{New Challenges}} for {{Justice}} and {{Accountability}}}.
\newblock \bibinfo{journal}{\emph{Annual Review of Law and Social Science}} \bibinfo{volume}{16}, \bibinfo{number}{1} (\bibinfo{date}{Oct.} \bibinfo{year}{2020}), \bibinfo{pages}{223--240}.
\newblock
\showISSN{1550-3585, 1550-3631}
\href{https://doi.org/10.1146/annurev-lawsocsci-060220-081955}{doi:\nolinkurl{10.1146/annurev-lawsocsci-060220-081955}}


\bibitem[Liu et~al\mbox{.}(2022)]%
        {liuRelationalMemoryAugmentedLanguage2022}
\bibfield{author}{\bibinfo{person}{Qi Liu}, \bibinfo{person}{Dani Yogatama}, {and} \bibinfo{person}{Phil Blunsom}.} \bibinfo{year}{2022}\natexlab{}.
\newblock \showarticletitle{Relational {{Memory-Augmented Language Models}}}.
\newblock \bibinfo{journal}{\emph{Transactions of the Association for Computational Linguistics}}  \bibinfo{volume}{10} (\bibinfo{date}{May} \bibinfo{year}{2022}), \bibinfo{pages}{555--572}.
\newblock
\showISSN{2307-387X}
\href{https://doi.org/10.1162/tacl_a_00476}{doi:\nolinkurl{10.1162/tacl_a_00476}}


\bibitem[Luca(2023)]%
        {EPRS_New_Product_Liability_Directive_2023}
\bibfield{author}{\bibinfo{person}{Stefano~De Luca}.} \bibinfo{year}{2023}\natexlab{}.
\newblock \bibinfo{booktitle}{\emph{New Product Liability Directive}}.
\newblock \bibinfo{type}{Briefing} PE 739.341. \bibinfo{institution}{European Parliamentary Research Service}.
\newblock
\urldef\tempurl%
\url{https://eur-lex.europa.eu/legal-content/EN/TXT/?uri=celex:52022PC0495}
\showURL{%
\tempurl}
\newblock
\shownote{EU Legislation in Progress}.


\bibitem[Lucas(2024)]%
        {lucasTeensMakingFriends2024}
\bibfield{author}{\bibinfo{person}{Jessica Lucas}.} \bibinfo{year}{2024}\natexlab{}.
\newblock \bibinfo{title}{The Teens Making Friends with {{AI}} Chatbots --- Theverge.Com}.
\newblock


\bibitem[Luger(2016)]%
        {lugerCHIHavingReally2016c}
\bibfield{author}{\bibinfo{person}{Abigail Luger, Ewa;~Sellen}.} \bibinfo{year}{2016}\natexlab{}.
\newblock \bibinfo{title}{{{CHI}} - "like Having a Really Bad {{PA}}": {{The}} Gulf between User Expectation and Experience of Conversational Agents}.
\newblock \bibinfo{numpages}{5286--5297}~pages.
\newblock
\href{https://doi.org/10.1145/2858036.2858288}{doi:\nolinkurl{10.1145/2858036.2858288}}


\bibitem[L{\"u}thi et~al\mbox{.}(2023)]%
        {luthiAugmentedIntelligenceAugmented2023}
\bibfield{author}{\bibinfo{person}{Nick L{\"u}thi}, \bibinfo{person}{Christian Matt}, \bibinfo{person}{Thomas Myrach}, {and} \bibinfo{person}{Iris Junglas}.} \bibinfo{year}{2023}\natexlab{}.
\newblock \showarticletitle{Augmented {{Intelligence}}, {{Augmented Responsibility}}?}
\newblock \bibinfo{journal}{\emph{Business \& Information Systems Engineering}} \bibinfo{volume}{65}, \bibinfo{number}{4} (\bibinfo{date}{Aug.} \bibinfo{year}{2023}), \bibinfo{pages}{391--401}.
\newblock
\showISSN{2363-7005, 1867-0202}
\href{https://doi.org/10.1007/s12599-023-00789-9}{doi:\nolinkurl{10.1007/s12599-023-00789-9}}


\bibitem[Ma et~al\mbox{.}(2023)]%
        {maUnderstandingBenefitsChallenges2023}
\bibfield{author}{\bibinfo{person}{Zilin Ma}, \bibinfo{person}{Yiyang Mei}, {and} \bibinfo{person}{Zhaoyuan Su}.} \bibinfo{year}{2023}\natexlab{}.
\newblock \bibinfo{title}{Understanding the {{Benefits}} and {{Challenges}} of {{Using Large Language Model-based Conversational Agents}} for {{Mental Well-being Support}}}.
\newblock
\showeprint[arxiv]{2307.15810}~[cs]
\href{https://doi.org/10.48550/arXiv.2307.15810}{doi:\nolinkurl{10.48550/arXiv.2307.15810}}


\bibitem[Meng et~al\mbox{.}(2023)]%
        {mengMediatedSocialSupport2023}
\bibfield{author}{\bibinfo{person}{Jingbo Meng}, \bibinfo{person}{Minjin~(MJ) Rheu}, \bibinfo{person}{Yue Zhang}, \bibinfo{person}{Yue Dai}, {and} \bibinfo{person}{Wei Peng}.} \bibinfo{year}{2023}\natexlab{}.
\newblock \showarticletitle{Mediated Social Support for Distress Reduction: {{AI}} Chatbots vs. Human}.
\newblock \bibinfo{journal}{\emph{Proc. ACM Hum.-Comput. Interact.}} \bibinfo{volume}{7}, \bibinfo{number}{CSCW1}, Article \bibinfo{articleno}{72} (\bibinfo{date}{April} \bibinfo{year}{2023}).
\newblock
\href{https://doi.org/10.1145/3579505}{doi:\nolinkurl{10.1145/3579505}}


\bibitem[Metz(2020)]%
        {metzRidingOutQuarantine2020}
\bibfield{author}{\bibinfo{person}{Cade Metz}.} \bibinfo{year}{2020}\natexlab{}.
\newblock \showarticletitle{Riding {{Out Quarantine With}} a {{Chatbot Friend}}: `{{I Feel Very Connected}}'}.
\newblock \bibinfo{journal}{\emph{The New York Times}} (\bibinfo{date}{June} \bibinfo{year}{2020}).
\newblock
\showISSN{0362-4331}


\bibitem[{Muhammad Azam} et~al\mbox{.}(2024)]%
        {muhammadazamEnhancingChatbotIntelligence2024}
\bibfield{author}{\bibinfo{person}{{Muhammad Azam}}, \bibinfo{person}{{Lubaina Zafar}}, \bibinfo{person}{{Tanveer Rafiq}}, \bibinfo{person}{Sana Zafar}, \bibinfo{person}{Umar Rafiq}, {and} \bibinfo{person}{Muhammad Adnan}.} \bibinfo{year}{2024}\natexlab{}.
\newblock \showarticletitle{Enhancing {{Chatbot Intelligence Through Narrative Memory Structures}}}.
\newblock \bibinfo{journal}{\emph{The Asian Bulletin of Big Data Management}} \bibinfo{volume}{4}, \bibinfo{number}{02} (\bibinfo{date}{May} \bibinfo{year}{2024}).
\newblock
\showISSN{2959-0809, 2959-0795}
\href{https://doi.org/10.62019/abbdm.v4i02.154}{doi:\nolinkurl{10.62019/abbdm.v4i02.154}}


\bibitem[Muresan and Pohl(2019)]%
        {muresanChatsBotsBalancing2019a}
\bibfield{author}{\bibinfo{person}{Andreea Muresan} {and} \bibinfo{person}{Henning Pohl}.} \bibinfo{year}{2019}\natexlab{}.
\newblock \showarticletitle{Chats with Bots: {{Balancing}} Imitation and Engagement}. In \bibinfo{booktitle}{\emph{Extended Abstracts of the 2019 {{CHI}} Conference on Human Factors in Computing Systems}} \emph{(\bibinfo{series}{Chi Ea '19})}. \bibinfo{publisher}{Association for Computing Machinery}, \bibinfo{address}{New York, NY, USA}, \bibinfo{pages}{1--6}.
\newblock
\showISBNx{978-1-4503-5971-9}
\href{https://doi.org/10.1145/3290607.3313084}{doi:\nolinkurl{10.1145/3290607.3313084}}


\bibitem[Namvarpour and Razi(2024)]%
        {namvarpour2024UncoveringContradictionsHumanAI}
\bibfield{author}{\bibinfo{person}{Mohammad Namvarpour} {and} \bibinfo{person}{Afsaneh Razi}.} \bibinfo{year}{2024}\natexlab{}.
\newblock \showarticletitle{Uncovering {{Contradictions}} in {{Human-AI Interactions}}: {{Lessons Learned}} from {{User Reviews}} of {{Replika}}}. In \bibinfo{booktitle}{\emph{Companion {{Publication}} of the 2024 {{Conference}} on {{Computer-Supported Cooperative Work}} and {{Social Computing}}}}. \bibinfo{publisher}{ACM}, \bibinfo{address}{San Jose Costa Rica}, \bibinfo{pages}{579--586}.
\newblock
\showISBNx{979-8-4007-1114-5}
\href{https://doi.org/10.1145/3678884.3681909}{doi:\nolinkurl{10.1145/3678884.3681909}}


\bibitem[Namvarpour and Razi(2025)]%
        {10.1145/3719160.3736621}
\bibfield{author}{\bibinfo{person}{Mohammad~(Matt) Namvarpour} {and} \bibinfo{person}{Afsaneh Razi}.} \bibinfo{year}{2025}\natexlab{}.
\newblock \showarticletitle{The Art of Talking Machines: A Comprehensive Literature Review of Conversational User Interfaces}. In \bibinfo{booktitle}{\emph{Proceedings of the 7th ACM Conference on Conversational User Interfaces}} \emph{(\bibinfo{series}{CUI '25})}. \bibinfo{publisher}{Association for Computing Machinery}, \bibinfo{address}{New York, NY, USA}, Article \bibinfo{articleno}{45}, \bibinfo{numpages}{18}~pages.
\newblock
\showISBNx{9798400715273}
\href{https://doi.org/10.1145/3719160.3736621}{doi:\nolinkurl{10.1145/3719160.3736621}}


\bibitem[Nasim et~al\mbox{.}(2022)]%
        {nasimArtificialIntelligenceIncidents2022}
\bibfield{author}{\bibinfo{person}{Syeda~Faiza Nasim}, \bibinfo{person}{{Muhammad Rizwan Ali}}, {and} \bibinfo{person}{{Umme Kulsoom}}.} \bibinfo{year}{2022}\natexlab{}.
\newblock \showarticletitle{Artificial {{Intelligence Incidents}} \& {{Ethics A Narrative Review}}}.
\newblock \bibinfo{journal}{\emph{International Journal of Technology, Innovation and Management (IJTIM)}} \bibinfo{volume}{2}, \bibinfo{number}{2} (\bibinfo{date}{Oct.} \bibinfo{year}{2022}).
\newblock
\showISSN{2789-777X}
\href{https://doi.org/10.54489/ijtim.v2i2.80}{doi:\nolinkurl{10.54489/ijtim.v2i2.80}}


\bibitem[Neff and Nagy(2016)]%
        {neffTalkingBotsSymbiotic2016}
\bibfield{author}{\bibinfo{person}{Gina Neff} {and} \bibinfo{person}{Peter Nagy}.} \bibinfo{year}{2016}\natexlab{}.
\newblock \showarticletitle{Talking to Bots: {{Symbiotic}} Agency and the Case of Tay}.
\newblock \bibinfo{journal}{\emph{International Journal of Communication}}  \bibinfo{volume}{10} (\bibinfo{date}{Oct.} \bibinfo{year}{2016}), \bibinfo{pages}{4915--4931}.
\newblock


\bibitem[Norman(2013)]%
        {normanDesignEverydayThings2013}
\bibfield{author}{\bibinfo{person}{Donald~A. Norman}.} \bibinfo{year}{2013}\natexlab{}.
\newblock \bibinfo{booktitle}{\emph{The Design of Everyday Things} (\bibinfo{edition}{revised and expanded edition} ed.)}.
\newblock \bibinfo{publisher}{Basic Books}, \bibinfo{address}{New York, New York}.
\newblock
\showISBNx{978-0-465-05065-9}
\showLCCN{TS171.4 .N67 2013}


\bibitem[of~the European~Communities(1985)]%
        {EU_Product_Liability_Directive_1985}
\bibfield{author}{\bibinfo{person}{Council of~the European~Communities}.} \bibinfo{year}{1985}\natexlab{}.
\newblock \bibinfo{title}{Council Directive 85/374/EEC of 25 July 1985 on the approximation of the laws, regulations and administrative provisions of the Member States concerning liability for defective products}.
\newblock \bibinfo{numpages}{29--33}~pages.
\newblock
\urldef\tempurl%
\url{https://eur-lex.europa.eu/legal-content/EN/TXT/?uri=CELEX%3A31985L0374}
\showURL{%
\tempurl}
\newblock
\shownote{Finnish special edition: Chapter 15 Volume 6 P. 0239; Spanish special edition: Chapter 13 Volume 19 P. 0008; Swedish special edition: Chapter 15 Volume 6 P. 0239; Portuguese special edition: Chapter 13 Volume 19 P. 0008}.


\bibitem[OpenAI(2023a)]%
        {openaiGPT35turbo2023}
\bibfield{author}{\bibinfo{person}{OpenAI}.} \bibinfo{year}{2023}\natexlab{a}.
\newblock \bibinfo{title}{{{GPT-3}}.5-Turbo}.
\newblock


\bibitem[OpenAI(2023b)]%
        {openaiGPT42023}
\bibfield{author}{\bibinfo{person}{OpenAI}.} \bibinfo{year}{2023}\natexlab{b}.
\newblock \bibinfo{title}{{{GPT-4}}}.
\newblock


\bibitem[Osiris1953(2021)]%
        {osiris1953UpdateRegardingGPT2021}
\bibfield{author}{\bibinfo{person}{Osiris1953}.} \bibinfo{year}{2021}\natexlab{}.
\newblock \bibinfo{title}{Update {{Regarding GPT}}}.
\newblock


\bibitem[Park et~al\mbox{.}(2021)]%
        {parkUseOffensiveLanguage2021}
\bibfield{author}{\bibinfo{person}{Namkee Park}, \bibinfo{person}{Kyungeun Jang}, \bibinfo{person}{Seonggyeol Cho}, {and} \bibinfo{person}{Jinyoung Choi}.} \bibinfo{year}{2021}\natexlab{}.
\newblock \showarticletitle{Use of Offensive Language in Human-Artificial Intelligence Chatbot Interaction: {{The}} Effects of Ethical Ideology, Social Competence, and Perceived Humanlikeness}.
\newblock \bibinfo{journal}{\emph{Computers in Human Behavior}}  \bibinfo{volume}{121} (\bibinfo{date}{Aug.} \bibinfo{year}{2021}), \bibinfo{pages}{106795}.
\newblock
\showISSN{0747-5632}
\href{https://doi.org/10.1016/j.chb.2021.106795}{doi:\nolinkurl{10.1016/j.chb.2021.106795}}


\bibitem[Parliament and Council(2024)]%
        {EU_AI_Act_2024}
\bibfield{author}{\bibinfo{person}{European Parliament} {and} \bibinfo{person}{Council}.} \bibinfo{year}{2024}\natexlab{}.
\newblock \bibinfo{title}{Regulation (EU) 2024/1689 of the European Parliament and of the Council of 13 June 2024 laying down harmonised rules on artificial intelligence and amending various regulations and directives}.
\newblock
\urldef\tempurl%
\url{http://data.europa.eu/eli/reg/2024/1689/oj}
\showURL{%
\tempurl}
\newblock
\shownote{Official Journal of the European Union, L 1689, 12.7.2024}.


\bibitem[Pentina et~al\mbox{.}(2023)]%
        {pentinaExploringRelationshipDevelopment2023}
\bibfield{author}{\bibinfo{person}{Iryna Pentina}, \bibinfo{person}{Tyler Hancock}, {and} \bibinfo{person}{Tianling Xie}.} \bibinfo{year}{2023}\natexlab{}.
\newblock \showarticletitle{Exploring Relationship Development with Social Chatbots: {{A}} Mixed-Method Study of Replika}.
\newblock \bibinfo{journal}{\emph{Computers in Human Behavior}}  \bibinfo{volume}{140} (\bibinfo{date}{March} \bibinfo{year}{2023}), \bibinfo{pages}{107600}.
\newblock
\showISSN{0747-5632}
\href{https://doi.org/10.1016/j.chb.2022.107600}{doi:\nolinkurl{10.1016/j.chb.2022.107600}}


\bibitem[Purington et~al\mbox{.}(2017)]%
        {puringtonAlexaMyNew2017a}
\bibfield{author}{\bibinfo{person}{Amanda Purington}, \bibinfo{person}{Jessie~G. Taft}, \bibinfo{person}{Shruti Sannon}, \bibinfo{person}{Natalya~N. Bazarova}, {and} \bibinfo{person}{Samuel~Hardman Taylor}.} \bibinfo{year}{2017}\natexlab{}.
\newblock \showarticletitle{"{{Alexa}} Is My New {{BFF}}": {{Social Roles}}, {{User Satisfaction}}, and {{Personification}} of the {{Amazon Echo}}}. In \bibinfo{booktitle}{\emph{Proceedings of the 2017 {{CHI Conference Extended Abstracts}} on {{Human Factors}} in {{Computing Systems}}}}. \bibinfo{publisher}{{ACM}}, \bibinfo{address}{{Denver Colorado USA}}, \bibinfo{pages}{2853--2859}.
\newblock
\showISBNx{978-1-4503-4656-6}
\href{https://doi.org/10.1145/3027063.3053246}{doi:\nolinkurl{10.1145/3027063.3053246}}


\bibitem[Razi et~al\mbox{.}(2023)]%
        {razi2023sliding}
\bibfield{author}{\bibinfo{person}{Afsaneh Razi}, \bibinfo{person}{Ashwaq Alsoubai}, \bibinfo{person}{Seunghyun Kim}, \bibinfo{person}{Shiza Ali}, \bibinfo{person}{Gianluca Stringhini}, \bibinfo{person}{Munmun De~Choudhury}, {and} \bibinfo{person}{Pamela~J. Wisniewski}.} \bibinfo{year}{2023}\natexlab{}.
\newblock \showarticletitle{Sliding into My DMs: Detecting Uncomfortable or Unsafe Sexual Risk Experiences within Instagram Direct Messages Grounded in the Perspective of Youth}.
\newblock \bibinfo{journal}{\emph{Proc. ACM Hum.-Comput. Interact.}} \bibinfo{volume}{7}, \bibinfo{number}{CSCW1}, Article \bibinfo{articleno}{89} (\bibinfo{date}{April} \bibinfo{year}{2023}), \bibinfo{numpages}{29}~pages.
\newblock
\href{https://doi.org/10.1145/3579522}{doi:\nolinkurl{10.1145/3579522}}


\bibitem[Razi et~al\mbox{.}(2020)]%
        {razi2020let}
\bibfield{author}{\bibinfo{person}{Afsaneh Razi}, \bibinfo{person}{Karla Badillo-Urquiola}, {and} \bibinfo{person}{Pamela~J. Wisniewski}.} \bibinfo{year}{2020}\natexlab{}.
\newblock \showarticletitle{Let's Talk about Sext: How Adolescents Seek Support and Advice about Their Online Sexual Experiences}. In \bibinfo{booktitle}{\emph{Proceedings of the 2020 CHI Conference on Human Factors in Computing Systems}} (Honolulu, HI, USA) \emph{(\bibinfo{series}{CHI '20})}. \bibinfo{publisher}{Association for Computing Machinery}, \bibinfo{address}{New York, NY, USA}, \bibinfo{pages}{1–13}.
\newblock
\showISBNx{9781450367080}
\href{https://doi.org/10.1145/3313831.3376400}{doi:\nolinkurl{10.1145/3313831.3376400}}


\bibitem[Razi et~al\mbox{.}(2021)]%
        {razi2021human}
\bibfield{author}{\bibinfo{person}{Afsaneh Razi}, \bibinfo{person}{Seunghyun Kim}, \bibinfo{person}{Ashwaq Alsoubai}, \bibinfo{person}{Gianluca Stringhini}, \bibinfo{person}{Thamar Solorio}, \bibinfo{person}{Munmun De~Choudhury}, {and} \bibinfo{person}{Pamela~J. Wisniewski}.} \bibinfo{year}{2021}\natexlab{}.
\newblock \showarticletitle{A Human-Centered Systematic Literature Review of the Computational Approaches for Online Sexual Risk Detection}.
\newblock \bibinfo{journal}{\emph{Proc. ACM Hum.-Comput. Interact.}} \bibinfo{volume}{5}, \bibinfo{number}{CSCW2}, Article \bibinfo{articleno}{465} (\bibinfo{date}{Oct.} \bibinfo{year}{2021}), \bibinfo{numpages}{38}~pages.
\newblock
\href{https://doi.org/10.1145/3479609}{doi:\nolinkurl{10.1145/3479609}}


\bibitem[Razi et~al\mbox{.}(2024)]%
        {razi2024trauma}
\bibfield{author}{\bibinfo{person}{Afsaneh Razi}, \bibinfo{person}{John~S. Seberger}, \bibinfo{person}{Ashwaq Alsoubai}, \bibinfo{person}{Nurun Naher}, \bibinfo{person}{Munmun De~Choudhury}, {and} \bibinfo{person}{Pamela~J. Wisniewski}.} \bibinfo{year}{2024}\natexlab{}.
\newblock \showarticletitle{Toward Trauma-Informed Research Practices with Youth in HCI: Caring for Participants and Research Assistants When Studying Sensitive Topics}.
\newblock \bibinfo{journal}{\emph{Proc. ACM Hum.-Comput. Interact.}} \bibinfo{volume}{8}, \bibinfo{number}{CSCW1}, Article \bibinfo{articleno}{134} (\bibinfo{date}{April} \bibinfo{year}{2024}), \bibinfo{numpages}{31}~pages.
\newblock
\href{https://doi.org/10.1145/3637411}{doi:\nolinkurl{10.1145/3637411}}


\bibitem[Reiss(2023)]%
        {reissTestingReliabilityChatGPT2023}
\bibfield{author}{\bibinfo{person}{Michael~V. Reiss}.} \bibinfo{year}{2023}\natexlab{}.
\newblock \bibinfo{title}{Testing the {{Reliability}} of {{ChatGPT}} for {{Text Annotation}} and {{Classification}}: {{A Cautionary Remark}}}.
\newblock
\showeprint[arxiv]{2304.11085}~[cs]


\bibitem[Rosemary et~al\mbox{.}(2024)]%
        {rosemaryRELATIONSHIPANONYMITYCYBER2024}
\bibfield{author}{\bibinfo{person}{Rizanna Rosemary}, \bibinfo{person}{Arief~Bayu Wardhana}, \bibinfo{person}{Hamdani~M. Syam}, {and} \bibinfo{person}{Novi Susilawati}.} \bibinfo{year}{2024}\natexlab{}.
\newblock \showarticletitle{{{THE RELATIONSHIP BETWEEN ANONYMITY AND CYBER SEXUAL HARASSMENT BY TWITTER USERS}}: {{A CROSS-SECTIONAL STUDY}}}.
\newblock \bibinfo{journal}{\emph{Journal of Community Mental Health and Public Policy}} \bibinfo{volume}{6}, \bibinfo{number}{2} (\bibinfo{date}{April} \bibinfo{year}{2024}), \bibinfo{pages}{95--104}.
\newblock
\showISSN{2622-2655}
\href{https://doi.org/10.51602/cmhp.v6i2.131}{doi:\nolinkurl{10.51602/cmhp.v6i2.131}}


\bibitem[Samuel(2024)]%
        {samuelPeopleAreFalling2024}
\bibfield{author}{\bibinfo{person}{Sigal Samuel}.} \bibinfo{year}{2024}\natexlab{}.
\newblock \showarticletitle{People Are Falling in Love with - and Getting Addicted to - {{AI}} Voices}.
\newblock \bibinfo{journal}{\emph{Vox}} (\bibinfo{date}{Aug.} \bibinfo{year}{2024}).
\newblock


\bibitem[Sapiro(2018)]%
        {sapiroSexualHarassmentPerformances2018}
\bibfield{author}{\bibinfo{person}{Virginia Sapiro}.} \bibinfo{year}{2018}\natexlab{}.
\newblock \showarticletitle{Sexual {{Harassment}}: {{Performances}} of {{Gender}}, {{Sexuality}}, and {{Power}}}.
\newblock \bibinfo{journal}{\emph{Perspectives on Politics}} \bibinfo{volume}{16}, \bibinfo{number}{4} (\bibinfo{date}{Dec.} \bibinfo{year}{2018}), \bibinfo{pages}{1053--1066}.
\newblock
\showISSN{1537-5927, 1541-0986}
\href{https://doi.org/10.1017/S1537592718002815}{doi:\nolinkurl{10.1017/S1537592718002815}}


\bibitem[Seberger et~al\mbox{.}(2022)]%
        {sebergerStillCreepyAll2022}
\bibfield{author}{\bibinfo{person}{John~S. Seberger}, \bibinfo{person}{Irina Shklovski}, \bibinfo{person}{Emily Swiatek}, {and} \bibinfo{person}{Sameer Patil}.} \bibinfo{year}{2022}\natexlab{}.
\newblock \showarticletitle{Still {{Creepy After All These Years}}:{{The Normalization}} of {{Affective Discomfort}} in {{App Use}}}. In \bibinfo{booktitle}{\emph{{{CHI Conference}} on {{Human Factors}} in {{Computing Systems}}}}. \bibinfo{publisher}{ACM}, \bibinfo{address}{New Orleans LA USA}, \bibinfo{pages}{1--19}.
\newblock
\showISBNx{978-1-4503-9157-3}
\href{https://doi.org/10.1145/3491102.3502112}{doi:\nolinkurl{10.1145/3491102.3502112}}


\bibitem[Skjuve et~al\mbox{.}(2022)]%
        {skjuveLongitudinalStudyHuman2022}
\bibfield{author}{\bibinfo{person}{Marita Skjuve}, \bibinfo{person}{Asbj{\o}rn F{\o}lstad}, \bibinfo{person}{Knut~Inge Fostervold}, {and} \bibinfo{person}{Petter~Bae Brandtzaeg}.} \bibinfo{year}{2022}\natexlab{}.
\newblock \showarticletitle{A Longitudinal Study of Human{\textendash}Chatbot Relationships}.
\newblock \bibinfo{journal}{\emph{International Journal of Human-Computer Studies}}  \bibinfo{volume}{168} (\bibinfo{date}{Dec.} \bibinfo{year}{2022}), \bibinfo{pages}{102903}.
\newblock
\showISSN{1071-5819}
\href{https://doi.org/10.1016/j.ijhcs.2022.102903}{doi:\nolinkurl{10.1016/j.ijhcs.2022.102903}}


\bibitem[Smith(2023)]%
        {smithIfCanPredict2023}
\bibfield{author}{\bibinfo{person}{Phoebe Smith}.} \bibinfo{year}{2023}\natexlab{}.
\newblock \showarticletitle{If {{I Can}}'t {{Predict My Future}}, {{Why Can AI}}? {{Exploring Human Interaction}} with {{Predictive Analytics}}}.
\newblock \bibinfo{journal}{\emph{Theses - ALL}} (\bibinfo{date}{May} \bibinfo{year}{2023}).
\newblock


\bibitem[St{\aa}hl and Dennhag(2021)]%
        {stahlOnlineOfflineSexual2021}
\bibfield{author}{\bibinfo{person}{Simon St{\aa}hl} {and} \bibinfo{person}{Inga Dennhag}.} \bibinfo{year}{2021}\natexlab{}.
\newblock \showarticletitle{Online and Offline Sexual Harassment Associations of Anxiety and Depression in an Adolescent Sample}.
\newblock \bibinfo{journal}{\emph{Nordic Journal of Psychiatry}} \bibinfo{volume}{75}, \bibinfo{number}{5} (\bibinfo{date}{July} \bibinfo{year}{2021}), \bibinfo{pages}{330--335}.
\newblock
\showISSN{0803-9488, 1502-4725}
\href{https://doi.org/10.1080/08039488.2020.1856924}{doi:\nolinkurl{10.1080/08039488.2020.1856924}}


\bibitem[Stephanie(2014)]%
        {stephanieCohenKappaStatistic2014}
\bibfield{author}{\bibinfo{person}{Stephanie}.} \bibinfo{year}{2014}\natexlab{}.
\newblock \bibinfo{title}{Cohen's {{Kappa Statistic}}}.
\newblock \bibinfo{howpublished}{https://www.statisticshowto.com/cohens-kappa-statistic/}.
\newblock


\bibitem[Strengers et~al\mbox{.}(2021)]%
        {strengersWhatCanHCI2021}
\bibfield{author}{\bibinfo{person}{Yolande Strengers}, \bibinfo{person}{Jathan Sadowski}, \bibinfo{person}{Zhuying Li}, \bibinfo{person}{Anna Shimshak}, {and} \bibinfo{person}{Florian 'Floyd'~Mueller}.} \bibinfo{year}{2021}\natexlab{}.
\newblock \showarticletitle{What {{Can HCI Learn}} from {{Sexual Consent}}? {{A Feminist Process}} of {{Embodied Consent}} for {{Interactions}} with {{Emerging Technologies}}}. In \bibinfo{booktitle}{\emph{Proceedings of the 2021 {{CHI Conference}} on {{Human Factors}} in {{Computing Systems}}}} \emph{(\bibinfo{series}{{{CHI}} '21})}. \bibinfo{publisher}{{Association for Computing Machinery}}, \bibinfo{address}{{New York, NY, USA}}, \bibinfo{pages}{1--13}.
\newblock
\showISBNx{978-1-4503-8096-6}
\href{https://doi.org/10.1145/3411764.3445107}{doi:\nolinkurl{10.1145/3411764.3445107}}


\bibitem[Str{\"o}mel et~al\mbox{.}(2024)]%
        {stromelNarratingFitnessLeveraging2024}
\bibfield{author}{\bibinfo{person}{Konstantin~R. Str{\"o}mel}, \bibinfo{person}{Stanislas Henry}, \bibinfo{person}{Tim Johansson}, \bibinfo{person}{Jasmin Niess}, {and} \bibinfo{person}{Pawe{\l}~W. Wo{\'z}niak}.} \bibinfo{year}{2024}\natexlab{}.
\newblock \showarticletitle{Narrating Fitness: {{Leveraging}} Large Language Models for Reflective Fitness Tracker Data Interpretation}. In \bibinfo{booktitle}{\emph{Proceedings of the 2024 {{CHI}} Conference on Human Factors in Computing Systems}} \emph{(\bibinfo{series}{Chi '24})}. \bibinfo{publisher}{Association for Computing Machinery}, \bibinfo{address}{New York, NY, USA}, Article \bibinfo{articleno}{646}.
\newblock
\showISBNx{9798400703300}
\href{https://doi.org/10.1145/3613904.3642032}{doi:\nolinkurl{10.1145/3613904.3642032}}


\bibitem[Subiyantoro et~al\mbox{.}(2023)]%
        {subiyantoroExploringImpactAIPowered2023}
\bibfield{author}{\bibinfo{person}{Singgih Subiyantoro}, \bibinfo{person}{I~Nyoman~Sudana Degeng}, \bibinfo{person}{Dedi Kuswandi}, {and} \bibinfo{person}{Saida Ulfa}.} \bibinfo{year}{2023}\natexlab{}.
\newblock \showarticletitle{Exploring the {{Impact}} of {{AI-Powered Chatbots}} ({{Chat GPT}}) on {{Education}}: {{A Qualitative Study}} on {{Benefits}} and {{Drawbacks}}}.
\newblock \bibinfo{journal}{\emph{Jurnal Pekommas}} \bibinfo{volume}{8}, \bibinfo{number}{2} (\bibinfo{date}{Dec.} \bibinfo{year}{2023}), \bibinfo{pages}{157--168}.
\newblock
\showISSN{2502-1907, 2502-1893}
\href{https://doi.org/10.56873/jpkm.v8i2.5206}{doi:\nolinkurl{10.56873/jpkm.v8i2.5206}}


\bibitem[Svikhnushina et~al\mbox{.}(2021)]%
        {svikhnushinaUserExpectationsConversational2021}
\bibfield{author}{\bibinfo{person}{Ekaterina Svikhnushina}, \bibinfo{person}{Alexandru Placinta}, {and} \bibinfo{person}{Pearl Pu}.} \bibinfo{year}{2021}\natexlab{}.
\newblock \showarticletitle{User {{Expectations}} of {{Conversational Chatbots Based}} on {{Online Reviews}}}. In \bibinfo{booktitle}{\emph{Designing {{Interactive Systems Conference}} 2021}} \emph{(\bibinfo{series}{{{DIS}} '21})}. \bibinfo{publisher}{{Association for Computing Machinery}}, \bibinfo{address}{{New York, NY, USA}}, \bibinfo{pages}{1481--1491}.
\newblock
\showISBNx{978-1-4503-8476-6}
\href{https://doi.org/10.1145/3461778.3462125}{doi:\nolinkurl{10.1145/3461778.3462125}}


\bibitem[{Ta-Johnson} et~al\mbox{.}(2022)]%
        {ta-johnsonAssessingTopicsMotivating2022}
\bibfield{author}{\bibinfo{person}{Vivian~P. {Ta-Johnson}}, \bibinfo{person}{Carolynn Boatfield}, \bibinfo{person}{Xinyu Wang}, \bibinfo{person}{Esther DeCero}, \bibinfo{person}{Isabel~C. Krupica}, \bibinfo{person}{Sophie~D. Rasof}, \bibinfo{person}{Amelie Motzer}, {and} \bibinfo{person}{Wiktoria~M. Pedryc}.} \bibinfo{year}{2022}\natexlab{}.
\newblock \showarticletitle{Assessing the {{Topics}} and {{Motivating Factors Behind Human-Social Chatbot Interactions}}: {{Thematic Analysis}} of {{User Experiences}}}.
\newblock \bibinfo{journal}{\emph{JMIR Human Factors}} \bibinfo{volume}{9}, \bibinfo{number}{4} (\bibinfo{date}{Oct.} \bibinfo{year}{2022}), \bibinfo{pages}{e38876}.
\newblock
\href{https://doi.org/10.2196/38876}{doi:\nolinkurl{10.2196/38876}}


\bibitem[Tranberg(2023)]%
        {tranbergLoveMyAI2023}
\bibfield{author}{\bibinfo{person}{Caroline Tranberg}.} \bibinfo{year}{2023}\natexlab{}.
\newblock \emph{\bibinfo{title}{``{{I}} Love My {{AI}} Girlfriend'' {{A}} Study of Consent in {{AI-human}} Relationships.}}
\newblock \bibinfo{thesistype}{Master's\ thesis}. \bibinfo{school}{The University of Bergen}.
\newblock


\bibitem[Trothen(2022)]%
        {trothenReplikaSpiritualEnhancement2022}
\bibfield{author}{\bibinfo{person}{Tracy~J. Trothen}.} \bibinfo{year}{2022}\natexlab{}.
\newblock \showarticletitle{Replika: {{Spiritual Enhancement Technology}}?}
\newblock \bibinfo{journal}{\emph{Religions}} \bibinfo{volume}{13}, \bibinfo{number}{4} (\bibinfo{date}{April} \bibinfo{year}{2022}), \bibinfo{pages}{275}.
\newblock
\showISSN{2077-1444}
\href{https://doi.org/10.3390/rel13040275}{doi:\nolinkurl{10.3390/rel13040275}}


\bibitem[Vanden~Abeele(2021)]%
        {vandenabeeleDigitalWellbeingDynamic2021}
\bibfield{author}{\bibinfo{person}{Mariek M~P Vanden~Abeele}.} \bibinfo{year}{2021}\natexlab{}.
\newblock \showarticletitle{Digital {{Wellbeing}} as a {{Dynamic Construct}}}.
\newblock \bibinfo{journal}{\emph{Communication Theory}} \bibinfo{volume}{31}, \bibinfo{number}{4} (\bibinfo{date}{Nov.} \bibinfo{year}{2021}), \bibinfo{pages}{932--955}.
\newblock
\showISSN{1050-3293, 1468-2885}
\href{https://doi.org/10.1093/ct/qtaa024}{doi:\nolinkurl{10.1093/ct/qtaa024}}


\bibitem[Varon and Pe{\~n}a(2021)]%
        {varonArtificialIntelligenceConsent2021}
\bibfield{author}{\bibinfo{person}{Joana Varon} {and} \bibinfo{person}{Paz Pe{\~n}a}.} \bibinfo{year}{2021}\natexlab{}.
\newblock \showarticletitle{Artificial Intelligence and Consent: A Feminist Anti-Colonial Critique}.
\newblock \bibinfo{journal}{\emph{Internet Policy Review}} \bibinfo{volume}{10}, \bibinfo{number}{4} (\bibinfo{date}{Dec.} \bibinfo{year}{2021}).
\newblock
\showISSN{2197-6775}
\href{https://doi.org/10.14763/2021.4.1602}{doi:\nolinkurl{10.14763/2021.4.1602}}


\bibitem[{Venkatesh} et~al\mbox{.}(2003)]%
        {venkateshUserAcceptanceInformation2003}
\bibfield{author}{\bibinfo{person}{{Venkatesh}}, \bibinfo{person}{{Morris}}, \bibinfo{person}{{Davis}}, {and} \bibinfo{person}{{Davis}}.} \bibinfo{year}{2003}\natexlab{}.
\newblock \showarticletitle{User {{Acceptance}} of {{Information Technology}}: {{Toward}} a {{Unified View}}}.
\newblock \bibinfo{journal}{\emph{MIS Quarterly}} \bibinfo{volume}{27}, \bibinfo{number}{3} (\bibinfo{year}{2003}), \bibinfo{pages}{425}.
\newblock
\showISSN{02767783}
\showeprint[jstor]{10.2307/30036540}
\href{https://doi.org/10.2307/30036540}{doi:\nolinkurl{10.2307/30036540}}


\bibitem[Vitak et~al\mbox{.}(2017)]%
        {vitakIdentifyingWomenExperiences2017}
\bibfield{author}{\bibinfo{person}{Jessica Vitak}, \bibinfo{person}{Kalyani Chadha}, \bibinfo{person}{Linda Steiner}, {and} \bibinfo{person}{Zahra Ashktorab}.} \bibinfo{year}{2017}\natexlab{}.
\newblock \showarticletitle{Identifying {{Women}}'s {{Experiences With}} and {{Strategies}} for {{Mitigating Negative Effects}} of {{Online Harassment}}}. In \bibinfo{booktitle}{\emph{Proceedings of the 2017 {{ACM Conference}} on {{Computer Supported Cooperative Work}} and {{Social Computing}}}}. \bibinfo{publisher}{ACM}, \bibinfo{address}{Portland Oregon USA}, \bibinfo{pages}{1231--1245}.
\newblock
\showISBNx{978-1-4503-4335-0}
\href{https://doi.org/10.1145/2998181.2998337}{doi:\nolinkurl{10.1145/2998181.2998337}}


\bibitem[Weidinger et~al\mbox{.}(2021)]%
        {weidingerEthicalSocialRisks2021a}
\bibfield{author}{\bibinfo{person}{Laura Weidinger}, \bibinfo{person}{John Mellor}, \bibinfo{person}{Maribeth Rauh}, \bibinfo{person}{Conor Griffin}, \bibinfo{person}{Jonathan Uesato}, \bibinfo{person}{Po-Sen Huang}, \bibinfo{person}{Myra Cheng}, \bibinfo{person}{Mia Glaese}, \bibinfo{person}{Borja Balle}, \bibinfo{person}{Atoosa Kasirzadeh}, \bibinfo{person}{Zac Kenton}, \bibinfo{person}{Sasha Brown}, \bibinfo{person}{Will Hawkins}, \bibinfo{person}{Tom Stepleton}, \bibinfo{person}{Courtney Biles}, \bibinfo{person}{Abeba Birhane}, \bibinfo{person}{Julia Haas}, \bibinfo{person}{Laura Rimell}, \bibinfo{person}{Lisa~Anne Hendricks}, \bibinfo{person}{William Isaac}, \bibinfo{person}{Sean Legassick}, \bibinfo{person}{Geoffrey Irving}, {and} \bibinfo{person}{Iason Gabriel}.} \bibinfo{year}{2021}\natexlab{}.
\newblock \showarticletitle{Ethical and Social Risks of Harm from {{Language Models}}}.
\newblock  (\bibinfo{year}{2021}).
\newblock
\href{https://doi.org/10.48550/ARXIV.2112.04359}{doi:\nolinkurl{10.48550/ARXIV.2112.04359}}


\bibitem[Xie and Pentina(2022)]%
        {xieAttachmentTheoryFramework2022}
\bibfield{author}{\bibinfo{person}{Tianling Xie} {and} \bibinfo{person}{Iryna Pentina}.} \bibinfo{year}{2022}\natexlab{}.
\newblock \bibinfo{booktitle}{\emph{Attachment {{Theory}} as a {{Framework}} to {{Understand Relationships}} with {{Social Chatbots}}: {{A Case Study}} of {{Replika}}}}.
\newblock
\showISBNx{978-0-9981331-5-7}


\bibitem[Young et~al\mbox{.}(2024)]%
        {young2024roleAI}
\bibfield{author}{\bibinfo{person}{Jordyn Young}, \bibinfo{person}{Laala~M Jawara}, \bibinfo{person}{Diep~N Nguyen}, \bibinfo{person}{Brian Daly}, \bibinfo{person}{Jina Huh-Yoo}, {and} \bibinfo{person}{Afsaneh Razi}.} \bibinfo{year}{2024}\natexlab{}.
\newblock \showarticletitle{The Role of AI in Peer Support for Young People: A Study of Preferences for Human- and AI-Generated Responses}. In \bibinfo{booktitle}{\emph{Proceedings of the 2024 CHI Conference on Human Factors in Computing Systems}} (Honolulu, HI, USA) \emph{(\bibinfo{series}{CHI '24})}. \bibinfo{publisher}{Association for Computing Machinery}, \bibinfo{address}{New York, NY, USA}, Article \bibinfo{articleno}{1006}, \bibinfo{numpages}{18}~pages.
\newblock
\showISBNx{9798400703300}
\href{https://doi.org/10.1145/3613904.3642574}{doi:\nolinkurl{10.1145/3613904.3642574}}


\bibitem[Zhang et~al\mbox{.}(2023)]%
        {zhangHowWouldStance2023}
\bibfield{author}{\bibinfo{person}{Bowen Zhang}, \bibinfo{person}{Daijun Ding}, {and} \bibinfo{person}{Liwen Jing}.} \bibinfo{year}{2023}\natexlab{}.
\newblock \bibinfo{title}{How Would {{Stance Detection Techniques Evolve}} after the {{Launch}} of {{ChatGPT}}?}
\newblock
\showeprint[arxiv]{2212.14548}~[cs]


\bibitem[Zimmerman et~al\mbox{.}(2023)]%
        {zimmermanHumanAIRelationships2023}
\bibfield{author}{\bibinfo{person}{Anne Zimmerman}, \bibinfo{person}{Joel Janhonen}, {and} \bibinfo{person}{Emily Beer}.} \bibinfo{year}{2023}\natexlab{}.
\newblock \showarticletitle{Human/{{AI}} Relationships: Challenges, Downsides, and Impacts on Human/Human Relationships}.
\newblock \bibinfo{journal}{\emph{AI and Ethics}} (\bibinfo{date}{Oct.} \bibinfo{year}{2023}).
\newblock
\showISSN{2730-5953, 2730-5961}
\href{https://doi.org/10.1007/s43681-023-00348-8}{doi:\nolinkurl{10.1007/s43681-023-00348-8}}


\end{thebibliography}

\end{document}